\documentclass{article}

\usepackage{arxiv}

\usepackage{amsmath,amsfonts}
\usepackage{algorithmic} % used instead of [noend] version
\usepackage{algorithm}
\usepackage{array}
\usepackage[caption=false,font=normalsize]{subfig} % preferred subcaption config
\usepackage{textcomp}
\usepackage{stfloats}
\usepackage{url}
\usepackage[hidelinks]{hyperref} % overrides plain hyperref
\usepackage{verbatim}
\usepackage{graphicx}
\usepackage{cite}
\usepackage{svg}
\usepackage{nicefrac}
\usepackage{microtype}
\usepackage{lipsum}
\usepackage{paracol}
\usepackage{caption}
\usepackage{amssymb}
\usepackage{float}
\usepackage{lineno}
\usepackage{siunitx}
\usepackage{booktabs}
\usepackage{mathtools}
\usepackage{xspace}
\usepackage{pifont}

\usepackage{needspace}
\usepackage{listings}
\usepackage{xcolor}

\newdimen\slantmathcorr
\def\oversl#1{%
  \setbox0=\hbox{$#1$}
  \slantmathcorr=\wd0
  \hskip 0.2\slantmathcorr \overline{\hbox to 0.8\wd0{%
  \vphantom{\hbox{$#1$}}}}
  \hskip-\wd0\hbox{$#1$}
}
\def\undersl#1{%
  \setbox0=\hbox{$#1$}
  \slantmathcorr=\wd0
  \underline{\hbox to 0.8\wd0{%
  \vphantom{\hbox{$#1$}}}}
  \hskip-0.8\wd0\hbox{$#1$}
}

\usepackage{amssymb}
\usepackage[table]{xcolor}
\definecolor{rowgray}{HTML}{E8E8E8}

\usepackage{needspace}
\usepackage{listings}
\usepackage{xcolor}

\usepackage{fancyhdr}

\begin{document}
\title{A General Approach to Visualizing Uncertainty in Statistical Graphics}

\author{
    Bernarda Petek \\
    University of Ljubljana\\
    Faculty of Computer and Information Science\\
    Večna pot 113, 1000 Ljubljana, Slovenia\\
    \texttt{bernarda.petek@fri.uni-lj.si} \\
\And
    David Nabergoj \\
    University of Ljubljana\\
    Faculty of Computer and Information Science\\
    Večna pot 113, 1000 Ljubljana, Slovenia\\
    \texttt{david.nabergoj@fri.uni-lj.si} \\
\And
Erik Štrumbelj \\
    University of Ljubljana\\
    Faculty of Computer and Information Science\\
    Večna pot 113, 1000 Ljubljana, Slovenia \\
    \texttt{erik.strumbelj@fri.uni-lj.si} \\
}

\date{}

\maketitle
\thispagestyle{fancy}

\begin{abstract}
We present a general approach to visualizing uncertainty in static 2-D statistical graphics. If we treat a visualization as a function of its underlying quantities, uncertainty in those quantities induces a distribution over images. We show how to aggregate these images into a single visualization that represents the uncertainty. The approach can be viewed as a generalization of sample-based approaches that use overlay. Notably, standard representations, such as confidence intervals and bands, emerge with their usual coverage guarantees without being explicitly quantified or visualized. As a proof of concept, we implement our approach in the IID setting using resampling, provided as an open-source Python library. Because the approach operates directly on images, the user needs only to supply the data and the code for visualizing the quantities of interest without uncertainty. Through several examples, we show how both familiar and novel forms of uncertainty visualization can be created. The implementation is not only a practical validation of the underlying theory but also an immediately usable tool that can complement existing uncertainty-visualization libraries.
\end{abstract}


\section{Introduction}

\begin{figure*}[t!]
    \centering
    \subfloat[univariate distribution]{%
        \includegraphics[width=.48\linewidth]{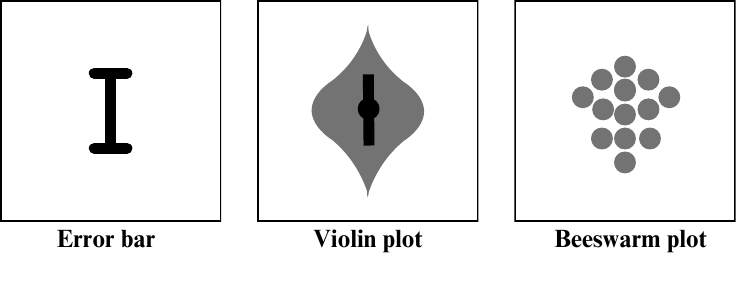}%
        \label{fig:examples:univariate}%
    }\hfill
    \subfloat[bivariate distribution]{%
        \includegraphics[width=.48\linewidth]{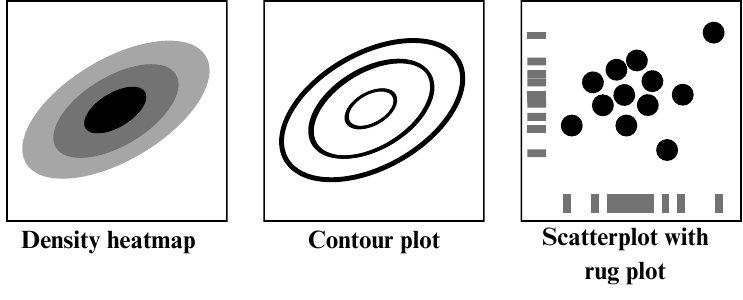}%
        \label{fig:examples:bivariate}%
    }\\
    \subfloat[distribution over functions]{%
        \includegraphics[width=.48\linewidth]{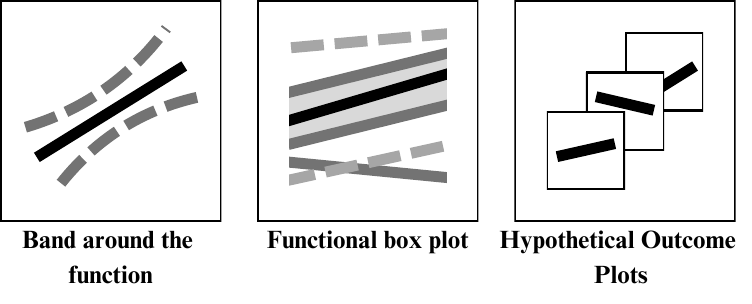}%
        \label{fig:examples:function}%
    }\hfill
    \subfloat[distribution over a probability simplex]{%
        \includegraphics[width=.48\linewidth]{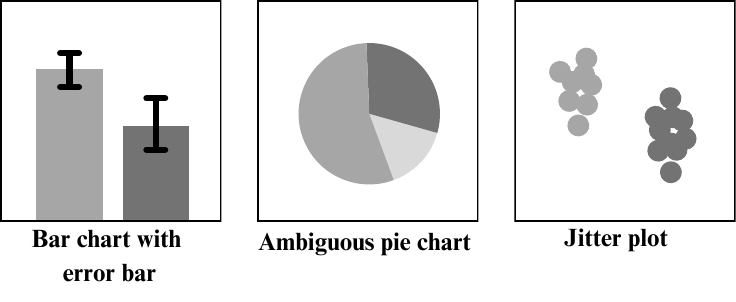}%
        \label{fig:examples:simplex}%
    }
    \caption{Illustrative examples of uncertainty visualizations in four common scenarios.}
    \label{fig:examples}
\end{figure*}

Communicating uncertainty is integral to modern quantitative data analysis, and visualization is the predominant mechanism for doing so (see Figure \ref{fig:examples} for illustrative examples). However, despite their importance, communicating uncertainty in general and uncertainty visualization in particular remain underutilized. This stems from both challenges in quantifying uncertainty and challenges in designing effective visual representations~\cite{hullman2019authors, panagiotidou2022communicating, fischhoff2012communicating}. With a lack of general approaches, an uncertainty visualization scenario typically requires a specific method for quantifying the uncertainty of interest and a specific method for visualizing it. This not only restricts most practitioners to uncertainty visualizations that are supported by popular software, but also leads to teaching-by-recipe and a limited understanding of uncertainty. 

In this paper, we propose a general approach for visualizing uncertainty in statistical graphics. The central idea is that a visualization of a quantity of interest, when treated without uncertainty, can be interpreted as a function applied to the distribution that represents uncertainty in that quantity. Evaluating this function over the underlying uncertainty distribution yields a distribution over images. For example, the code used to plot a sample mean can be viewed as a function of its sampling distribution. Applied to that distribution, it produces a corresponding distribution of mean plots. As demonstrated in Section~\ref{sec:method}, aggregating sample images drawn from this distribution produces a new visualization that directly represents the uncertainty in the quantity of interest.

Our approach simplifies the process of visualizing uncertainty both conceptually and in terms of practical application. It requires only two components: a distribution that represents our uncertainty (hereafter, \textbf{base distribution}) and the visualization that produces an image of the uncertain quantity of interest given one possible realization of its uncertainty (hereafter, \textbf{base visualization}). In other words, the user needs only to specify the base distribution and to plot the quantity of interest. Uncertainty of the quantity of interest is never explicitly computed or visualized. Instead, it is implicitly determined by how the quantity of interest is encoded in the base visualization and manifests in the final image through the variability of the base visualization outputs. For example, we do not have to compute the confidence interval (CI) from the base distribution or explicitly encode it in the visualization.

At first glance, the approach may appear to be just another sample-based method. However, the distinction lies in how the samples are incorporated. Sample-based techniques typically extend a base visualization by overlaying multiple sample realizations, animating them, or presenting them as small multiples. In contrast, our approach aggregates the images produced by the base visualization itself, treating that visualization as a black box and requiring no modification to how individual samples are rendered. Although the resulting visualizations may look similar when both approaches apply, the underlying principles differ in important ways. Notably, our method provides coverage guarantees that sample-based techniques have traditionally lacked. In this sense, our approach can be viewed as a generalization of sample-based methods that rely on overlayed samples. A more detailed discussion of how our approach relates to prior work is provided in Section~\ref{sec:related}.

To summarize, the main contributions of our work are:

\begin{itemize}
    \item A novel and general approach to visualizing uncertainty, offering a fundamentally different perspective on uncertainty in statistical graphics.
    \item Theoretical coverage guarantees for the resulting visualizations, which also illuminate properties of existing sample-based methods that rely on overlay.
    \item A reference implementation of the proposed approach for independent and identically distributed (IID) data, serving both as a versatile tool and as a foundation for developing new uncertainty visualizations.
\end{itemize}

\section{Background on Uncertainty Visualization}\label{sec:background}

The focus of our work is on communicating probabilistically describable uncertainty using static two-dimensional (2-D) visualizations, which remain the dominant form of uncertainty visualization.  For a comprehensive review of uncertainty visualization, we refer the reader to related work~\cite{bonneau2014overview,brodlie2012review,gillmann2021uncertainty,hagele2022uncertainty,jena2020uncertainty,kamal2021recent,maceachren2005visualizing,panagiotidou2022communicating,visUncBook,weiskopf2022uncertainty, hullman2018pursuit}.

We organize our review of static 2-D visualizations of uncertainty according to the domain of the base distribution and focus on the most popular approaches. This structure later facilitates discussion of the generality of our approach, which can generate suitable uncertainty visualizations for every common domain, as well as its limitations, since it cannot produce every type of visualization.

\subsection{Four Common Scenarios}

\subsubsection{Univariate Distribution}

The most common scenario involves an uncertain quantity represented by a single number (see Figure~\ref{fig:examples:univariate}), for example, the mean age. A standard approach is to represent uncertainty using a quantile-based interval, while the boxplot remains a popular alternative, particularly for empirical distributions. Entire univariate distributions can be visualized through their probability mass function by using bar charts or dot plots, or their probability density function, using density plots, histograms, or violin plots. Quantile dot plots, in contrast, simplify a probability distribution by discretizing it into evenly spaced quantile intervals and representing each interval with dots~\cite{kay2016ish}. Density strips offer a compact alternative, conveying density through gradients~\cite{jackson2008displaying}.

\subsubsection{Bivariate Distribution}

Another common scenario is that our uncertain quantity is a 2-D coordinate (see Figure~\ref{fig:examples:bivariate}), for example, a location on a map or the slope and intercept of our simple linear model. In 2-D density plots, uncertainty is typically represented with color (hue, intensity, gradient). A common alternative is a contour plot, where contour lines indicate regions of constant density. Instead of a smooth density transition, the data can be binned to produce a 2-D histogram or heatmap. There also exist many different versions of bivariate boxplots, which differ in what summary statistics are used, for example, the bagplot~\cite{rousseeuw1999bagplot}.

\subsubsection{Distribution Over Functions}

Another widely encountered scenario is when an uncertain quantity represents a functional relationship between two variables (see Figure~\ref{fig:examples:function}), for example, population growth over time or a dose-response curve. The most popular approaches are confidence bands and functional boxplots\cite{sun2011functional}.

\subsubsection{Distribution Over a Probability Simplex}

When our response of interest is a categorical variable but the probabilities of the categories are themselves uncertain, we have to visualize a distribution over a probability vector (see Figure~\ref{fig:examples:simplex}). With two and three categories, the probability simplex is a line segment and a triangle, respectively, so corresponding univariate and bivariate techniques can be used. With four or more categories, we have to resort to representing the probability and corresponding uncertainty of each category separately. Because probabilities are bounded, this is typically done with a bar chart or a dot plot (with a guideline) and with error bars representing uncertainty.

\subsection{Other Scenarios and Approaches}

The four scenarios described above encompass most cases of uncertainty visualization and the most commonly used approaches within them. Nevertheless, other techniques and less common, more complex scenarios also exist~\cite{brodlie2012review, potter2011quantification}. While we do not attempt a comprehensive review, Section~\ref{sec:examples} presents several uncertainty visualizations generated by our approach that extend beyond these scenarios, including representations of uncertainty in text, vectors, trajectories, flows, and multivariate data.

\subsection{Sample-Based Approaches}

Following the taxonomy in~\cite{weiskopf2022uncertainty}, the visualizations discussed above either represent uncertainty by explicitly showing the entire distribution or by visualizing summary statistics. The remaining options are to combine different visualization techniques or, as emphasized in this section, to visualize uncertainty implicitly via samples. We treat sample-based approaches separately for two reasons: (1) they can be applied across any of the previously discussed scenarios, and (2) they are closely related to our own work. The central idea of sample-based methods is to visualize multiple samples, typically ensuring that they are representative of the underlying distribution. Common strategies for combining these samples include overlay~\cite{gotz2019visualization}, animation~\cite{hullman2015hypothetical,kale2018hypothetical}, and small multiples~\cite{allen2019introduction,liu2018visualizing,mercier2019waterfall}.

Our approach uses overlay. While the concept of visualizing uncertainty by overlaying samples in a single visualization is not new, it has primarily been applied to specific cases. For univariate distributions, this typically results in variations of the scatterplot, such as strip plots, jitter plots, bee swarm plots, or rug plots. Functions or trajectories are often visualized as overlays, commonly referred to as line ensembles or spaghetti plots~\cite{padilla2020uncertainty}, including variants based on representative samples~\cite{liu2016uncertainty}. Similarly, a Multiple Forecast Visualization (MFV) chart displays multiple forecasts in the same visual space~\cite{padilla2022multiple}. Although the idea of generalizing overlay beyond these special cases may seem straightforward, it has not been systematically explored. It is implicitly suggested in~\cite{kale2018hypothetical} that such an overlay could be thought of as a static analogue of a Hypothetical Outcome Plot (HOP), but this notion is not further developed. We extend the concept of sample-based overlay methods by demonstrating that it can be generalized in several ways: (1) only the base visualization and distribution are required, eliminating the need to explicitly determine how samples are overlaid, (2) the resulting visualizations come with statistical guarantees, and (3) the range of useful uncertainty visualizations that can be produced is broad. Taken together, these generalizations are conceptually novel. In Sections~\ref{sec:related} and~\ref{sec:fuzzy_text}, we broaden the discussion by further comparing our approach to more specific approaches.

\section{Visualizing a Distribution over Statistical Graphics}\label{sec:method}

In this section, we define our approach to visualizing uncertainty. It requires only samples from the base distribution and a base visualization that produces an image of the quantity of interest for each sample. Combined, they result in a distribution over images that we aggregate into a single static 2-D visualization.

No further assumptions are made about the base distribution, making the approach agnostic to the method used to quantify uncertainty. Consequently, it can be applied to a wide range of outcomes, including those derived from Monte Carlo simulations, sensitivity analyses, classical statistical inference, and Bayesian inference. Likewise, we make no additional assumptions about the base visualization. Therefore, the approach can, in principle, be applied to any base visualization, treating it as a black box. In practice, however, not every base visualization will yield a useful visualization of uncertainty, a point we discuss in more detail in Sections~\ref{sec:bootplot} and~\ref{sec:examples}.

\subsection{A Visualization as a Statistical Functional}

First, we formalize the key terms of quantity of interest, base distribution, and base visualization. In a visualization of uncertainty, the\textbf{ quantity of interest} $\theta$ is uncertain and we define it to be a random variable. The uncertainty is represented by the \textbf{base distribution} of $\theta$, which we represent with its cumulative distribution function $F$ with support $\mathcal{S}$. Finally, the \textbf{base visualization} is a map $\mathcal{M}:  \mathcal{S} \rightarrow \{0, 1\}^{|\mathbb{R}|^2}$. That is, the sample is mapped to an image, where each point $(x, y) \in \mathbb{R} \times \mathbb{R}$ is a binary pixel. 

This definition of the base visualization is sufficient for our formal investigation of the approach. In practice, however, we must contend with color, finite pixel resolution, and the presence of multiple statistical functionals in a single visualization, to name just a few challenges. We address these practical issues in our reference implementation in Section~\ref{sec:bootplot}. Note also that it is immaterial whether the support $\mathcal{S}$ consists of numbers, coordinates, probability vectors, functions, or any other type of object, as long as there exists a base visualization that maps that object to an image.

When we combine the mapping $\mathcal{M}$ with the distribution $F$, we obtain a distribution over images, denoted $F_\mathcal{M}$. If we can sample $\theta$ from $F$ and apply $\mathcal{M}$ to that sample, we can thereby sample $\mathcal{M}(\theta)$ from the image distribution $F_\mathcal{M}$. Observe how the base visualization $\mathcal{M}$ is treated as a black box, so no understanding of its internal workings is required by the approach.

\begin{figure}[t!]
    \centering
    \subfloat[univariate]{%
        \includegraphics[width=\linewidth]{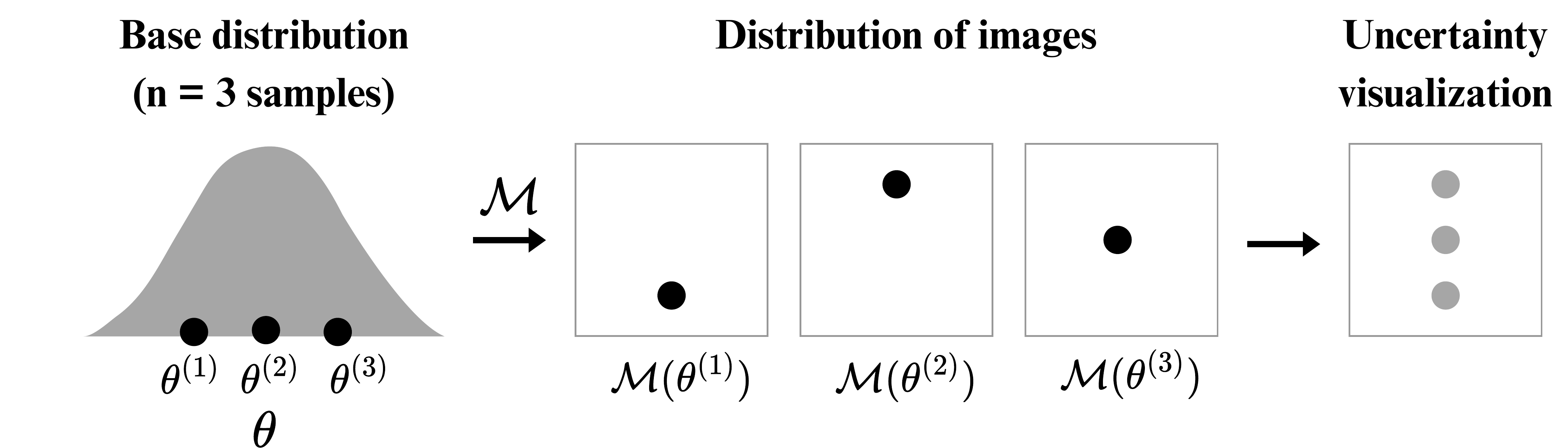}%
        \label{fig:point-example}%
    }\\[1.5ex]
    \subfloat[slope-intercept coordinate in simple regression]{%
        \includegraphics[width=\linewidth]{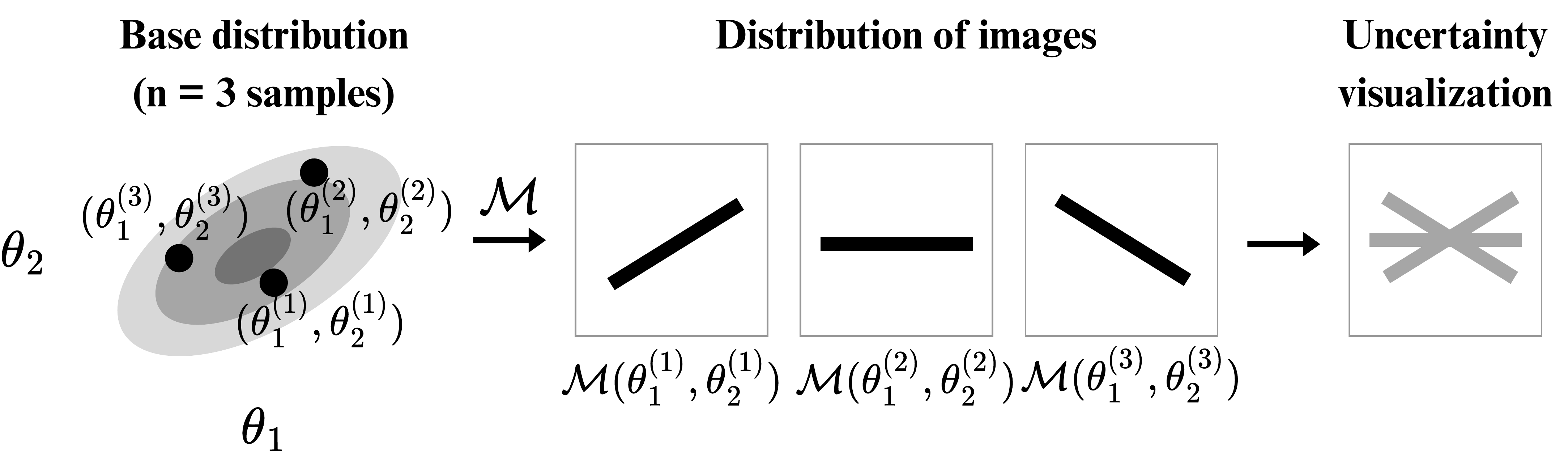}%
        \label{fig:line-example}%
    }
    \caption{Two examples of our approach to uncertainty visualization.}
    \label{fig:illustrative}
\end{figure}

Figures~\ref{fig:point-example} and~\ref{fig:line-example} illustrate the base distribution, base visualization, and the resulting distribution over images for two common practical scenarios. Note that in the case of linear regression, the quantity of interest is a coordinate of slope and intercept, but the base visualization is a line. That is, it is not uncommon that the visual representations are non-trivial transformations of the underlying quantity of interest.

\subsection{Visualizing a Distribution over Images}

We now have all the necessary components to define our approach to visualizing the uncertainty associated with a distribution over images $F_\mathcal{M}$. Given a sequence $M^{(1)}$, $M^{(2)}$ ,\dots, $M^{(n)}$ of independent samples from $F_\mathcal{M}$, we define a new image $M_{res}$ in a point-wise manner:

$$M_{res}(x, y) = \frac{1}{n} \sum_{i = 1}^n M^{(i)}(x, y), \mbox{ }\forall (x, y) \in \mathbb{R} \times \mathbb{R}.$$

We produce a new image by averaging the intensities of the sample images, for each pixel separately. The main advantage of this approach is that it maintains the black-box treatment of $\mathcal{M}$. While the approach is very simple, it is versatile and comes with formal guarantees in many standard uncertainty visualization tasks.

\subsection{Relationship between Sample Size and Coverage}\label{sec:coverage}

Figure~\ref{fig:illustrative} illustrates the approach on $n = 3$ samples. As $n$ increases, more extreme samples are produced. For a formal investigation, two cases are of interest: a set value of $n$ and as $n$ approaches infinity. We focus on the former, which is of more practical interest.

\subsubsection{A Point Varying in One Dimension}\label{sec:point}

First, we investigate the coverage of the interval observed in $M_{res}$ in the case where the map $\mathcal{M}$ is such that the realizations of $\theta$ always map to a single point and that point varies across $\theta$ along a single axis. An example would be the point estimation in Figure~\ref{fig:point-example}. That is, we limit ourselves to $\mathcal{M}$ where
$$\mathcal{M}(\theta)(x, y) = \begin{cases}
1, \text{for point } (f(\theta), 0)\\
0, \text{ otherwise},\\
\end{cases}$$

\noindent for some $f: \mathcal{S} \rightarrow \mathbb{R}$. Note that we assume, without loss of generality, that we vary along the $x$-axis. 

This simplifies the investigation to the univariate random variable $f(\theta)$. Let $X_1, X_2, \dots, X_n$ be independent samples from the same distribution. We further introduce the extreme observations and the range 
\begin{align*}
L_n &= \min\{X_1, X_2, \dots, X_n\},\\
U_n &= \max\{X_1, X_2, \dots, X_n\},\\
R_n &= [L_n, U_n].
\end{align*}

The range $R$ communicates the uncertainty associated with the random variable. The key question is, what is its coverage $P(X \in R_n | X_1,\dots, X_n)$, where X is an independent sample from the same distribution. 

The question falls into the theory of order statistics, but for our needs, a combinatorial argument will suffice: 
\begin{align*}
     & P(X \in R_n | X_1,\dots, X_n) = \\
     &= 1 - P(X \notin R_n | X_1,\dots, X_n) \\
     &= 1 - P(X > U_n| X_1,\dots, X_n) - P(X < L_n | X_1, \dots, X_n).
\end{align*}

There are $(n+1)!$ possible orderings of the $n + 1$ random variables ($X$ and the $n$ variables in the sample). If ties are not possible, there are $n!$ possible orderings where the $X$ is strictly the highest. If ties are possible, there are fewer than $n!$ such orderings. It follows that
$$P(X > U_n| X_1,\dots, X_n) \leq \frac{n!}{(n+1)!} = \frac{1}{n + 1}.$$

With a symmetric argument for the lower part, we get:
$$  P(X \in R_n | X_1,\dots, X_n) \geq 1 - \frac{2}{n + 1} = \frac{n-1}{n+1}.$$

Without ties, equality holds. Note that this result is independent of the distribution of $X$!

In the case of the visualization of a point that varies along an axis, this means that the coverage of the range formed by the two most extreme points that appeared in the images has coverage $\frac{n-1}{n+1}$, where $n$ is the number of images. Solving for a desired coverage $c$, we have the cutoff $n \geq \frac{c + 1}{1 - c}$, with $c \in [0, 1)$. Furthermore, commonly used coverages $80\%$, $90\%$, $95\%$, $99\%$, and $99.9\%$ all have integer solutions (see Table~\ref{tbl:coverages}). The practical implication is that the coverage required by the user can be achieved by setting the corresponding sample size $n$. This case of a single point varying along an axis also applies to collections of points that vary identically along an axis, for example, a bar in a bar plot.

\subsubsection{Points Varying in One Dimension}

Next, we investigate $M_{res}$ for a set of points that vary along parallel (but unique) axes. An example would be the regression lines in Figure \ref{fig:line-example}. Without loss of generality, we can assume that the axes are parallel to the $y$-axis. That is, each point has its own $x$-coordinate and the points vary in their $y$-coordinate depending on the value of $\theta$: 
$$\mathcal{M}(\theta)(x, y) = \begin{cases}
1, \text{ for points } (x, f_\theta(x)), x \in A \subset \mathbb{R}\\
0, \text{ otherwise},\\
\end{cases}$$

\noindent for some $f_\theta: A \rightarrow \mathbb{R}$. In essence, we are visualizing a random function.

Given a sequence $M^{(1)}$, $M^{(2)}$,\dots, $M^{(n)}$ of images and their corresponding functions $f_{\theta^{(1)}}$, $f_{\theta^{(2)}}$, \dots, $f_{\theta^{(n)}}$, define the functions
\begin{align*}
f_{\max}(x) &= \max\{f_{\theta^{(1)}}(x), f_{\theta^{(2)}}(x), \dots, f_{\theta^{(n)}}(x)\},\\
f_{\min}(x) &= \min\{f_{\theta^{(1)}}(x), f_{\theta^{(2)}}(x), \dots, f_{\theta^{(n)}}(x)\},  
\end{align*}

\noindent for all $x \in A$. These are the upper and lower bound functions for a sequence of $n$ images.

The coverage analysis for a single point from the previous section applies in a point-wise manner to a set of points. That is, the region bounded by the upper and lower bound functions $f_{\max}(x)$ and $f_{\min}(x)$ is a point-wise confidence band with coverage at least $\frac{n-1}{n+1}$. 

\subsubsection{Inference for a Predetermined Region} Imagine a region of the uncertainty visualization that is determined before observing the data. If we then observe that the graphical representation of the quantity of interest does not appear in that region in the aggregate image (and thus not in any of the sample images), then we can infer about the quantity of interest being in that region. Typical examples would be whether the quantity exceeds a threshold of interest.

Formally, let $R \subset \mathbb{R}^2$ be a region of the image that is independent of the sequence of images $M^{(1)}, \dots, M^{(n)}$. We introduce the indicator variables
$$I_i = \{\text{$M^{(i)}$ has no points with value 1 in region R}\}$$

\noindent and their sum $Z = \sum_{i = 1}^n I_i$.

$Z$ is \text{Binomial}$(n, p)$, with unknown probability $p$. Therefore, inference about the quantity of interest lying in the region reduces to inference about the proportion of a Binomial distribution.

We utilize Jeffreys interval, which is a Bayesian approach but also has strong frequentist properties~\cite{cai2005one}. Jeffreys interval is based on the Binomial likelihood and a $\text{Beta}(\frac{1}{2}, \frac{1}{2})$ prior. If we limit ourselves to the case $Z = n$, the $\alpha$-level interval is $[Q_\text{Beta}(n, \alpha), 1]$, where $Q_\text{Beta}(n, \alpha)$ is the quantile function of $\text{Beta}(n + \frac{1}{2}, \frac{1}{2})$ at $\alpha$.

Note that the theory does not require us to limit ourselves to the extreme case of $Z = n$, that no images has an observation in the region. The same process applies to any observed proportion, but accurately counting the number of observations in a region of the aggregate image is typically impossible.

\begin{table}[!t]
\centering
\caption{Samples' sizes and corresponding implied coverage.}
\label{tbl:coverages}
\begin{tabular}{|r|c|c|c|}
\hline
\raisebox{-3pt}{$n$} & \raisebox{-3pt}{$1 - \alpha = \frac{n - 1}{n + 1}$} & \raisebox{-3pt}{$\frac{n + 0.5}{n + 1}$}  & \raisebox{-3pt}{$Q_\text{Beta}(n, 0.05)$} \\[5pt]
\hline
9 & 0.800 & 0.9500 & 0.8126 \\
19 & 0.900 & 0.9750 & 0.9051 \\
39 & 0.950 & 0.9875 & 0.9522 \\
199 & 0.990 & 0.9975 & 0.9904 \\
1999 & 0.999 & 0.9998 & 0.9990 \\
\hline
\end{tabular}
\end{table}

Finally, we provide some coverage guarantees as a function of sample size $n$. Under the Binomial-Beta model with Jeffreys prior, the mean is $\frac{n + \frac{1}{2}}{n + 1}$, which converges from above to $\frac{n - 1}{n + 1}$, the coverage implied by $n$ (see Section~\ref{sec:point}). Furthermore, the Jeffreys interval lower bound $Q_\text{Beta}(n, 0.05)$ approximates well the implied coverage for that $n$ (see Table~\ref{tbl:coverages}).

The above results have the following practical implication. As demonstrated in Section~\ref{sec:point}, the number of samples $n$ can be chosen to achieve the desired coverage of observed confidence intervals and bands. We further show that this coverage also serves as a lower bound for inferences regarding whether the quantity of interest lies within a predetermined region. The practical application of these theoretical guarantees is  in Section~\ref{sec:examples_inference}.

\section{bootplot: A Reference Implementation}\label{sec:bootplot}

To further validate our approach and to demonstrate that it has not only theoretical but also practical utility, we provide a reference implementation. 

We limit the reference implementation to IID cases. However, this should not be interpreted as a limitation of the approach itself. In principle, the method can be trivially extended to more general sampling schemes, since it is agnostic to how the base distribution is generated. In practice, though, specifying and managing more complex sampling schemes significantly complicates the interface. This reflects the same fundamental challenge that prevents inference libraries from being both simple and general. A practical solution is to provide specialized libraries tailored to specific scenarios, such as generalized linear models, structural equation modeling, time series, etc.

The IID assumption simplifies obtaining the base distribution, as it allows the use of a straightforward resampling scheme: individual data points are sampled with replacement to create a resampled dataset of the same size, to which the base visualization is then applied. This procedure is analogous to a non-parametric bootstrap, with the key distinction that the functional of interest is an image rather than a scalar or vector. By leveraging this approach, we can focus on exploring the role of the base visualization while still addressing a wide range of practical scenarios.

\begin{figure}[!htbp]
   
    \begin{minipage}[t]{0.98\textwidth} % Adjust width to leave some gap
    \centering
     \begin{lstlisting}[]
# statistical graphic
def point_estimate_plot(data_subset, data_full, ax):
    ax.scatter(np.mean(data_subset), 0, color='black')
    ax.set(xlim=(-2, 4), ylim=(-1, 1))
    ax.get_figure().set_constrained_layout(True)

# call bootplot
bootplot(
    point_estimate_plot,
    dataset,
    m=39, # for 95% coverage
    output_image_path='point_estimate_plot.png',
    output_size_px=(900, 450)
)
\end{lstlisting}

    \end{minipage}
 \begin{minipage}[t]{0.5\linewidth} % [t] aligns tops, adjust width as needed
        \centering
\includegraphics[width=0.9\linewidth]{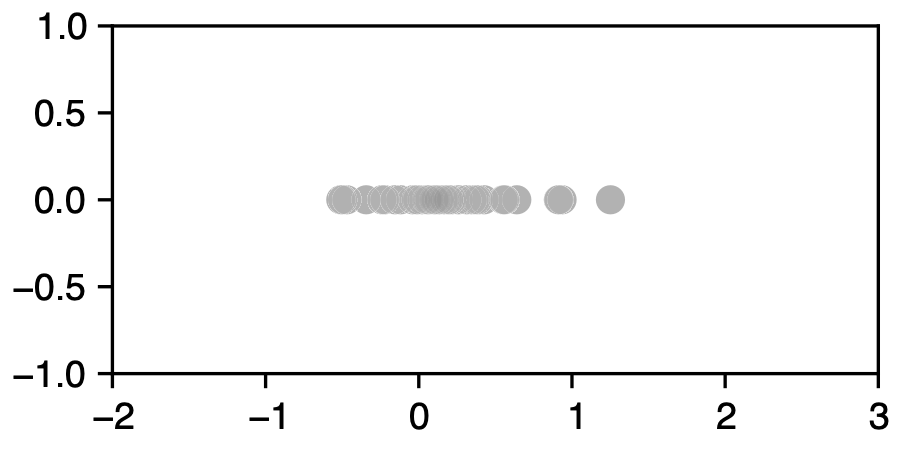}

\label{fig:simple_example}
    \end{minipage}%
      \begin{minipage}[t]{0.5\linewidth} % [t] aligns tops, adjust width as needed
      \centering
\includegraphics[width=0.9\linewidth]{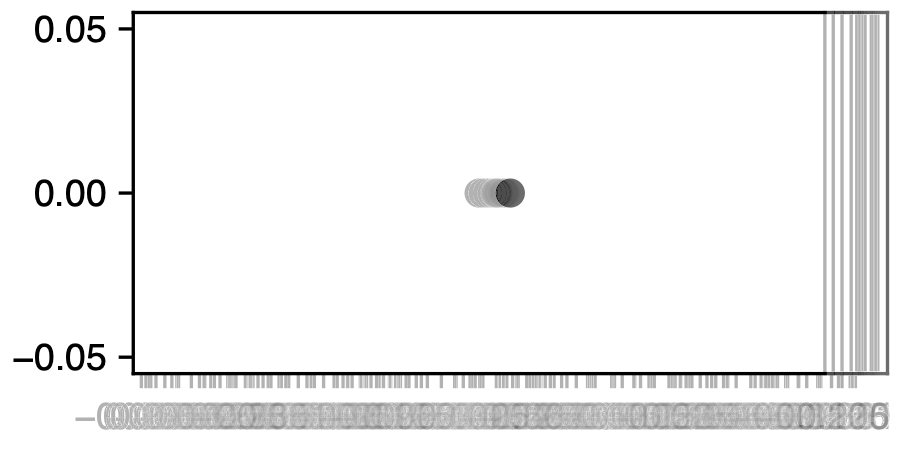}

    \end{minipage}
    \caption{This code snippet illustrates bootplot's use. The dataset is just an array of numbers, and the quantity of interest is the mean. The user provides code for visualizing the mean (the base visualization), following a prescribed signature, and passes it to a bootplot call, together with the rest of the parameters, including the number of samples, set to $39$ in this example. Both the left-hand and right-hand images are the result of the proposed method of generating images of multiple samples from the target distribution and then combining them. The only difference is that for the right-hand, the axes were not set (we removed line 4). Instead, the axes are set automatically and vary across samples, resulting in unwanted variability. The left-hand image shows 95\% confidence interval.} % Overall caption for the float
    \label{fig:bootplot}
\end{figure}

The reference implementation is available as a Python library (\url{https://github.com/davidnabergoj/bootplot}), with documentation and code for all examples that appear in the remainder of the paper. The user only has to provide the code for the base visualization, which takes data as an argument and visualizes the quantity of interest. Other parameters, including the number of samples $n$, have meaningful default values and do not need to be specified manually, but can be optionally adjusted by the user. Figure~\ref{fig:bootplot} shows example code with the corresponding output image.

\subsection{Practical Considerations and Limitations}

The implementation of the theoretical approach presents additional challenges due to how visualization is handled, both in Python and in computer graphics generally.

\subsubsection{Color}

We accommodate color by working in the RGB color space and aggregating each channel separately. This results in alpha blending. 

\subsubsection{Implied Intervals and Bands}

Due to the sample-based nature of the approach, the interior of the interval (band) might not be completely filled. This is particularly common when the variability is large and $n$ is small. Examples of this can be seen in the confidence bands on the left-hand side of Figure~\ref{fig:polynomial_linear_comparison} and in the first error bar in Figure~\ref{fig:bar_chart_comparison}. This does not invalidate the theoretical coverage guarantees, which still apply to the interval implied by the two most extreme observations. However, visible gaps may still be a perceptual limitation (see Section~\ref{subsec:futurework}).

\subsubsection{Anti-aliasing}

Anti-aliasing results in the quantity of interest being visualized with multiple and usually lighter shades along its border. As a result, the borders in the aggregate image may become less visible. In most cases, this is barely noticeable, but we recommend disabling anti-aliasing.

\subsubsection{Resolution}

The output image resolution matches that of the base visualization. Our approach does not require higher resolution than necessary to display the base visualization adequately. We adopt a default of 300 DPI, consistent with standard practices for high-quality, publication-ready figures. Note that the method is limited to creating raster images and is fundamentally incompatible with creating vector images, due to pixel-level aggregation.

\subsubsection{Intensity and Visibility}\label{sec:color:issue}

Averaging across images introduces a normalization issue that is not present in sampling-based estimation theory. If the visual representation has a small area (a point, a line), but the area over which it varies is relatively large, this can result in very low average intensity. This is not a theoretical issue, because the intensities will still represent the distribution. However, it is a practical issue due to the limitations of perceiving such low intensities. Untreated, it leads to a counter-intuitive phenomenon where increasing $n$ leads to less uncertainty in the final image. Note that some statistical graphics do not suffer from this issue, because the area representing the quantity of interest is always the same (pie chart, stacked bar chart, icon chart).

\begin{figure}[!t]
\centering
\includegraphics[width=0.6\linewidth]{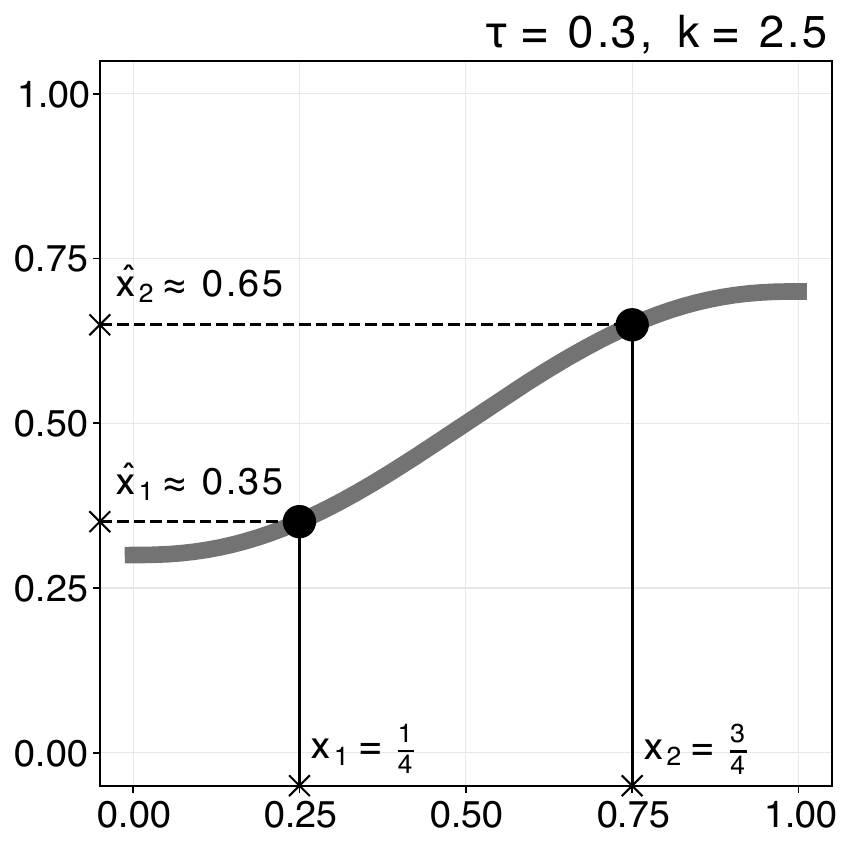}
\caption{An illustrative example of the transformation of the intensities in the case of two unique values.}
\label{fig:transformation_example}
\end{figure}

We deal with this issue by moving extreme intensities to a value where they can be perceived and interpolating the values in between to preserve the ordering. We transform the colors using a horizontally compressed CDF of the Beta distribution. We apply the same transformation to each RGB channel of each pixel, so we limit our discussion to one pixel and one channel. Given $n$ sample images, we have $n$ channel values for our pixel. If there is only one unique value that appears $n$ times, no transformation is necessary, as there is no variability and the channel value should be that unique value. Assume that there are $m \geq 2$ unique values and let $x = [x_1,\dots,x_m]$, $0 < x_i < 1$, be the probability vector of length $m$ representing the relative frequencies of the $m$ unique values. Our approach is based on adjusting these relative frequencies using the transformation

$$f(x, k, \tau) =  (1 - 2\tau) I_x(k, k) + \tau,$$

\noindent where $k > 0$ and $0 \leq \tau \leq 0.5$ are tunable parameters and $I_{x}(\alpha, \beta)$ is the regularized incomplete beta function (the CDF of the Beta distribution). See Figure~\ref{fig:transformation_example} for an illustrative example of this transformation.

We chose this CDF because it is monotonic from (0,0) to (1,1) and has a rotational symmetry about point $(\frac{1}{2}, \frac{1}{2})$. We require monotonicity because we are transforming relative frequencies into relative frequencies while preserving order. We require rotational symmetry, because the effect must be symmetrical for $x$ and $1 - x$. Furthermore, this CDF is flexible enough, with a single parameter, to express slow-then-fast ascent, fast-then-slow ascent, and linear ascent from $(0,0)$ to $(\frac{1}{2}, \frac{1}{2})$ (the other half from $(\frac{1}{2}, \frac{1}{2})$ to $(1,1)$ is determined by rotational symmetry).

First, we identify the largest frequency $c = \displaystyle\arg\max_i x_i$. If $c$ is not unique, we select it at random among the indices that maximize $x_i$. The transformed relative frequencies $\hat{x}_i$ are then  

$$
\hat{x}_i = \begin{cases}
  f(x_c, k, \tau)  & i = c \\
  f(1 - x_c, k, \tau) \frac{x_i}{1 - x_c} & i \neq c.
\end{cases}
$$

The transformation is most easily understood in the case of two unique values ($m = 2$), which is also the most common practical case, as we have the fixed background and the varying visualization of the quantity of interest (see Figure~\ref{fig:transformation_example} for an illustrative example). In the case of more than 2 unique values, we first transform the sum of the relative frequencies of all the values (except the maximal $c$) and then allocate the result proportionately to the original relative frequencies $x_i$. This ensures that the transformed relative frequencies are again a probability simplex.

\begin{figure}[!t]
\centering
    \centering
    % Row 1
    \subfloat[$\tau = 0$, $k = 1$]{\fbox{\includegraphics[width=0.3\linewidth,trim=230 170 250 220, clip]{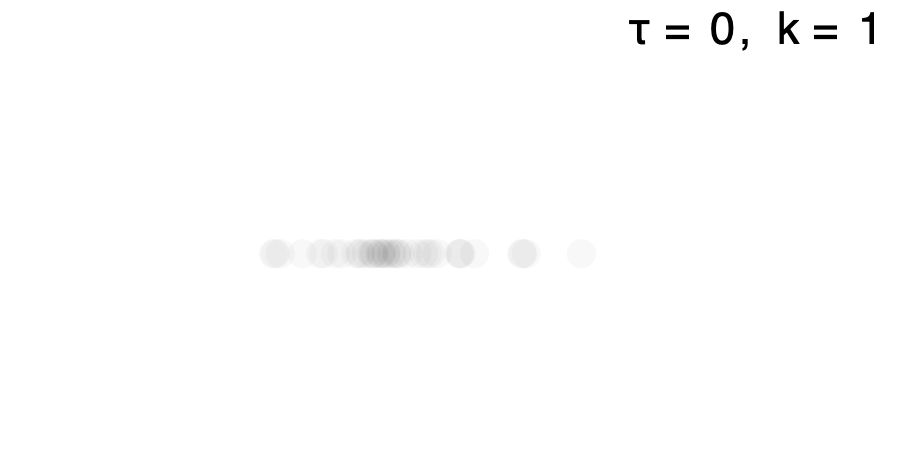}}} \hfill 
  \subfloat[$\tau = \frac{1}{10}$, $k = \frac{1}{2}$]{\fbox{\includegraphics[width=0.3\linewidth,trim=230 170 250 220, clip]{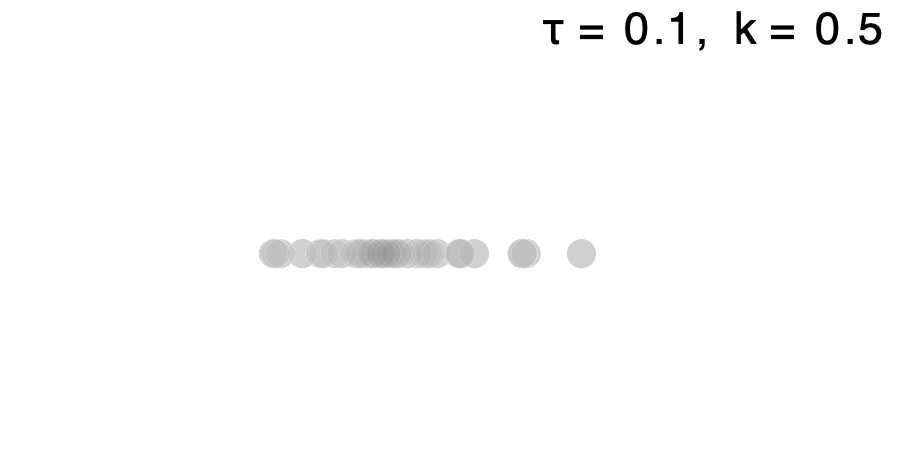}}}
    \hfill    
  \subfloat[$\tau = \frac{1}{10}$, $k = \frac{3}{2}$]{\fbox{\includegraphics[width=0.3\linewidth,trim=230 170 250 220, clip]{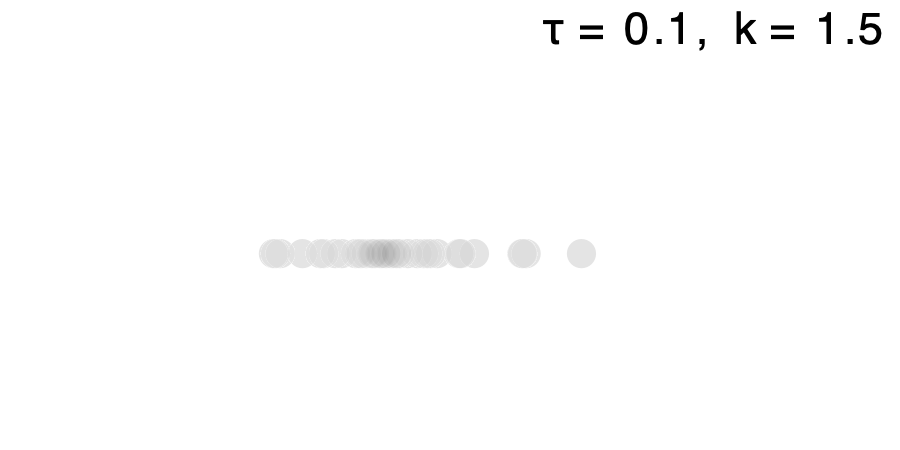}}}
    \hfill
    \\
    % Row 2
  \subfloat[$\tau = \frac{1}{10}$, $k = 10^6$]{\fbox{\includegraphics[width=0.3\linewidth,trim=230 170 250 220, clip]{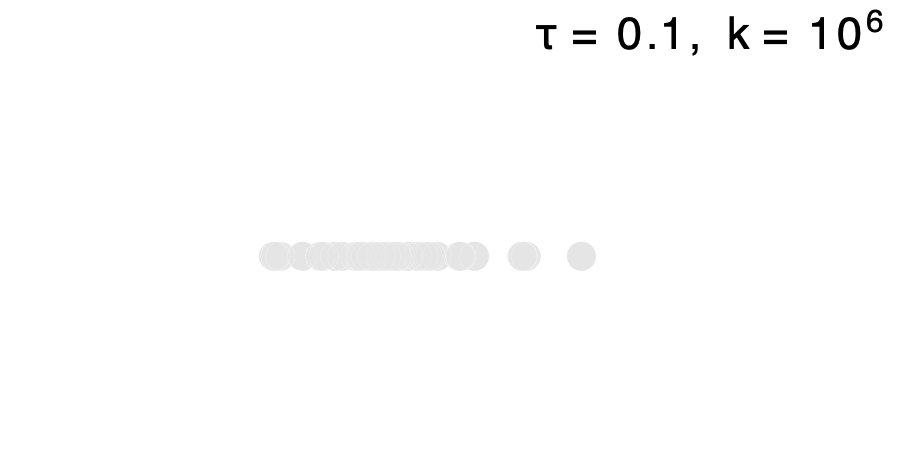}}}\hfill 
  \subfloat[$\tau = \frac{1}{2}$, $k = 1$]{\fbox{\includegraphics[width=0.3\linewidth,trim=230 170 250 220, clip]{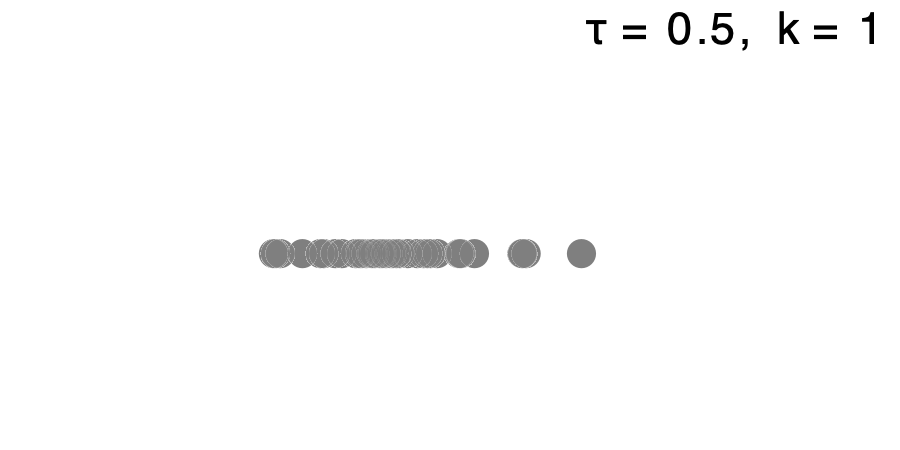}}}
    \hfill    
  \subfloat[$\tau = 0$, $k = 10^{-6}$]{\fbox{\includegraphics[width=0.3\linewidth,trim=230 170 250 220, clip]{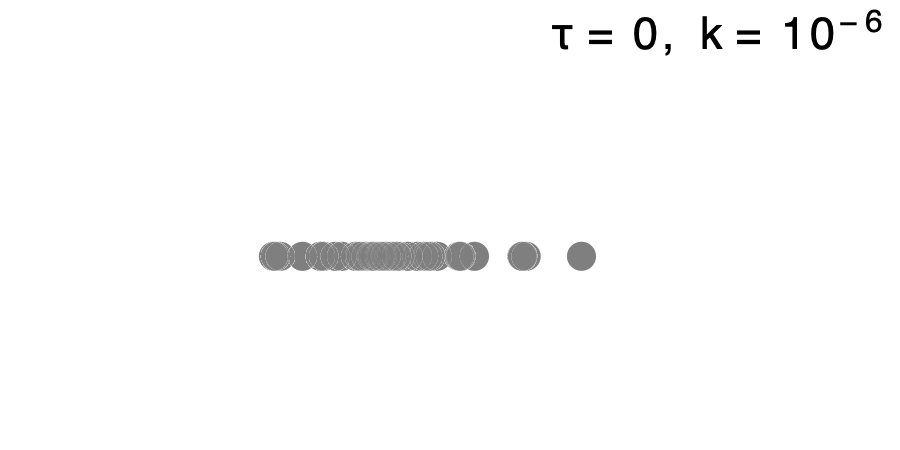}}}
    \hfill
    \\ 
\caption{The top-left image is the unaltered aggregate image ($\tau = 0$ and $k = 1$ is the identity transform). The bottom-middle image results in a uniform color ($\tau = \frac{1}{2}$ is a constant function). A uniform color also appears when the slope parameter $k$ is taken to either extreme, as shown in the bottom-left and bottom-right images. Intermediate values of $k$ control the gradient, as seen in the top-middle and top-right images.
}
\label{fig:parameters_choice}
\end{figure}

Users can adjust the threshold $\tau$ to improve visibility, and the slope $k$ to control the gradient of the visualized uncertainty (see Figure~\ref{fig:parameters_choice}). However, from our experience, adjustments are rarely necessary and all visualizations in this paper use the default settings of $\tau = 0.3$ and $k = 2.5$.

\subsubsection{Consistency of Layout}

Creating base visualizations requires careful attention to layout consistency. For example, if the axis bounds are determined automatically, each sample image may have different limits, resulting in misalignment and an unusable aggregate image (see Figure~\ref{fig:bootplot}). Common examples are axis bounds, axis tick positions and labels, order of variables, color mappings, and anchoring (first slice of pie chart, vectors in PCA/Factor Analysis; see Figure~\ref{fig:pie_chart_comparison} and Figure~\ref{fig:ex:pca}). To ensure consistent alignment of digits, a monospaced font should be used, and numbers should be aligned on the decimal point (see Figure~\ref{fig:polynomial_linear_comparison}). 

These considerations are rarely made when preparing a statistical visualization. Instead, we rely on the software to make these choices automatically. However, allowing such inconsistencies in our visualizations is poor practice. Automated choices can reduce semantic control, compromise reproducibility (since underlying software may change), and decrease transparency (because we do not know how the software made the choice). These issues most often go unnoticed because visualization code is rarely reused. However, when the same base visualization is applied repeatedly, as is the case in our approach, these issues become apparent. This is conceptually similar to the use of metamorphic testing for evaluating visualizations~\cite{mcnutt2020surfacing}, although in this work the approach primarily serves to visualize uncertainty.

In addition, because sampling from the base distribution can yield different visualizations for identical data, this process violates the invariance principle~\cite{kindlmann2014algebraic}. The resulting non-determinism in the visualization may negatively affect the user. To mitigate this issue, a fixed random seed should be specified.

\section{Examples}\label{sec:examples}

All examples use Matplotlib \cite{Hunter2007} for creating the base visualization. Most of the examples in this section use the Palmer penguins dataset~\cite{horst2020palmer}. Two examples use a simple dataset consisting of 59 air mass trajectories. All examples have $n = 39$ for $95\%$ coverage. We provide additional examples in the supplementary material illustrating the applicability and scalability of our approach to larger, real world datasets with more features and diverse underlying distributions. Additionally, while each example in this section is produced separately, our approach also enables users to visualize uncertainty in multiple statistical graphics simultaneously, which makes it easier to explore larger datasets. Importantly, examples in the supplementary material also show that larger datasets do not necessarily lead to fundamentally different visual outputs, but to greater certainty in the visualized quantities.

\subsection{Introductory Examples}

First, we reiterate the main properties of the proposed approach and further illustrate how the reference implementation is used. We illustrate this with two of the most common scenarios: visualizing uncertainty in a point estimate and in a functional relationship between two variables.

\begin{figure}[!t]
\centering
\includegraphics[width=3.5in]{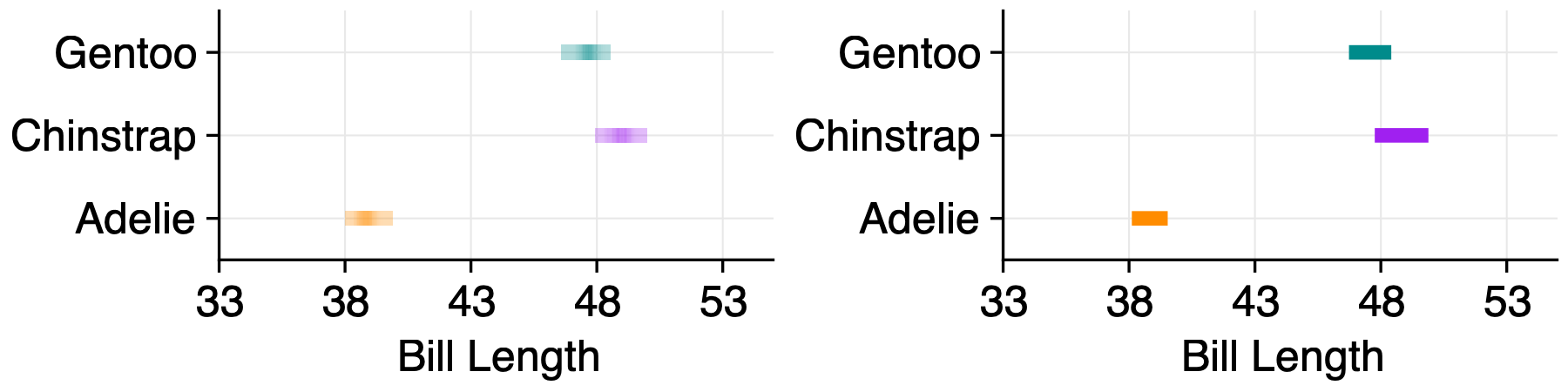}
\caption{Both images show a statistical graphic with 95\% confidence intervals for mean bill length for each species separately. The image on the left was created with bootplot having $n=39$ samples (the base visualization visualizes each mean with a point). The image on the right was created with Seaborn.}
\label{fig:points_ci_comparison}
\end{figure}

Figure~\ref{fig:points_ci_comparison} shows a typical visualization of uncertainty in a point estimate using CIs represented by an error bar. The image on the right was created with Seaborn \cite{Waskom2021}, a visualization library built on top of Matplotlib. Most of the elements of the visualization are specified in Matplotlib, while Seaborn provides support for visualizing 95\% CIs. Internally, Seaborn computes the CIs using bootstrap resampling and renders the corresponding error bars.

The image on the left was created with bootplot. The base visualization is a Matplotlib visualization that has its certain elements (see Section~\ref{sec:certain}) identical to the one used for Seaborn, except that each point estimate is visualized. This base visualization, the data, and the desired coverage are the only inputs the user must provide. The visualization is then combined in our approach with the base distribution, which is obtained by resampling the IID data. The number of resamples is uniquely determined by the desired coverage. Note that apart from the aesthetics and the presence of three estimates instead of one, this example is the same as the one in Figure \ref{fig:bootplot}.

Figure~\ref{fig:polynomial_linear_comparison} shows a typical visualization of uncertainty in a functional relationship. The image on the right is created using Seaborn, which computes the CIs and renders the corresponding band. The image on the left is created using bootplot. The base visualization is again identical in the certain parts to the one used for Seaborn, except that each regression fit is visualized as a function of the resampled data, as are the corresponding root mean squared errors (RMSE). This example demonstrates that the approach works with more than one quantity of interest in the same image, even when the graphical representations overlap. It also introduces the concept of uncertain text (the RMSE), which is discussed in more detail in Section~\ref{sec:fuzzy_text}

The intervals (bands) implied in the bootplot visualizations are similar to those in the Seaborn visualizations. This demonstrates the theoretical coverage guarantees of the approach, even though the intervals (bands) are never explicitly computed or explicitly visualized! Instead, they manifest in the aggregate image as a consequence of the uncertainty in the base visualization. It is important to note that these guarantees are not limited to these particular quantities of interest (mean, regression) or to the use of the bootstrap. They apply to any point varying along an axis, regardless of the statistical functional, the base distribution, or the method used to obtain it. Statistical inference using the visualizations created by our approach is discussed in Section~\ref{sec:examples_inference}.

\subsection{Certain and Uncertain Elements of the Base Visualization}\label{sec:certain}

\begin{figure}[!t]
\centering
\includegraphics[width=3.5in]{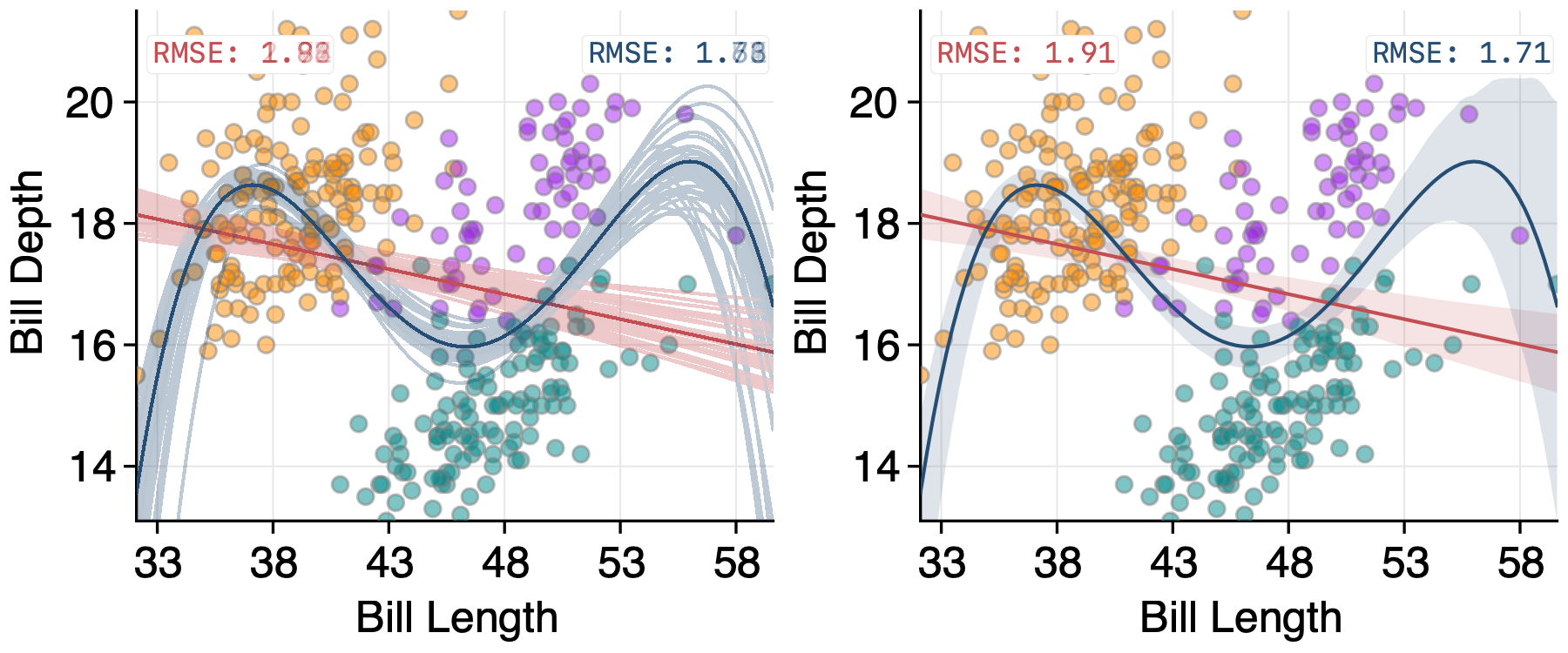}
\caption{Both images show a statistical graphic with 95\% point-wise confidence bands for the linear and polynomial regression fit of the relationship between bill length and bill depth. The image on the left was created with bootplot having $n=39$ samples (the base visualization visualizes the linear fit, the polynomial fit, and the RMSE of the two). The image on the right was created with Seaborn.}
\label{fig:polynomial_linear_comparison}
\end{figure}

In any base visualization, some visual elements, possibly none, are independent of the base distribution, while others, also possibly none, depend on it. For example, in the point estimates visualization shown in Figure~\ref{fig:points_ci_comparison}, all elements except the three point estimates of the mean (axes, labels, gridlines, etc.) are independent of the resampled data. Similarly, in the regression example in Figure~\ref{fig:polynomial_linear_comparison}, only the regression lines and the RMSE depend on the resampled data.

\begin{figure}[!t]
\centering
\includegraphics[width=3.5in]{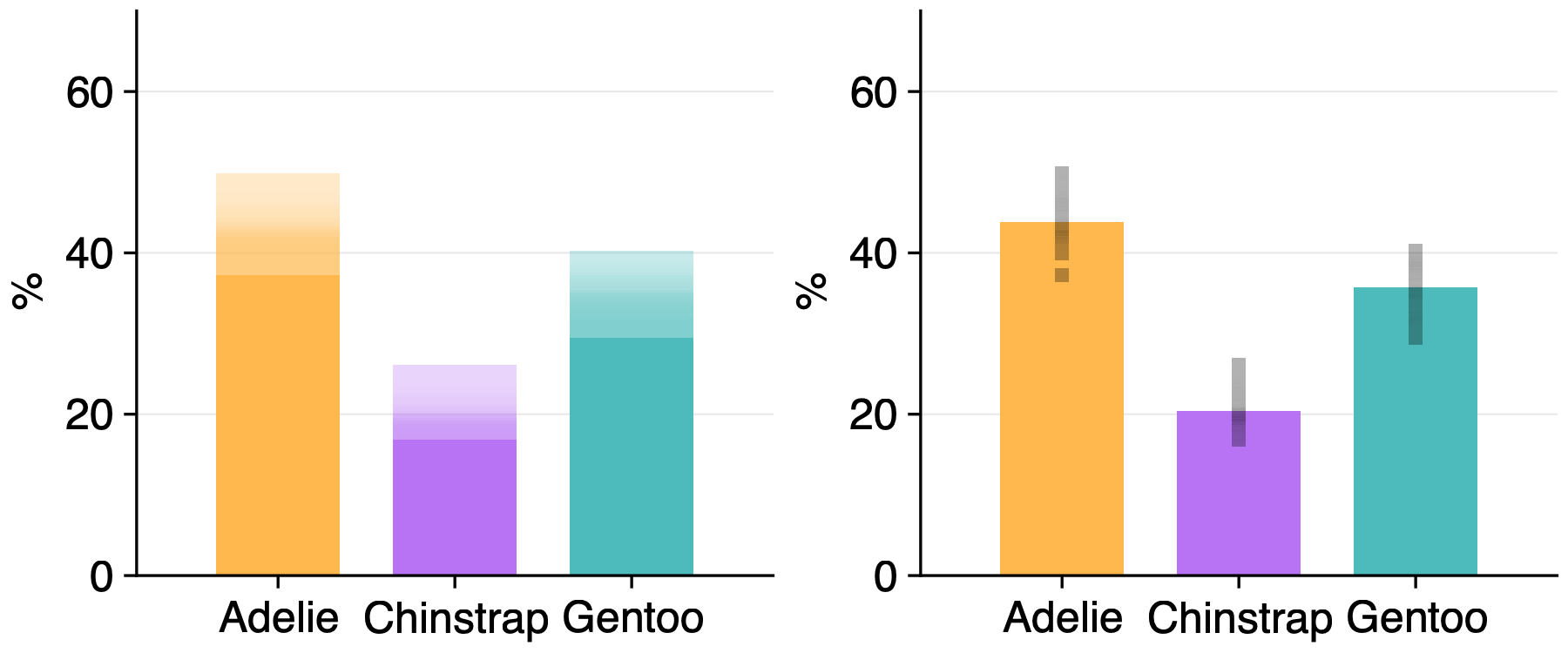}
\caption{When bootplot is applied to a base visualization of a bar chart, where the height of the bars is uncertain, representing proportions of penguin species, it results in an ambiguated bar chart~\cite{olston2002visualizing} (left-hand side image). If we instead keep the bars certain in the base visualization (a function of the data, not the resampled data) and add an uncertain point estimate to this base visualization, we create a bar chart with error bars (right-hand side image). Both images are created with bootplot having $n=39$ samples.}
\label{fig:bar_chart_comparison}
\end{figure}

\begin{figure}[!t]
\centering
\includegraphics[width=3.5in]{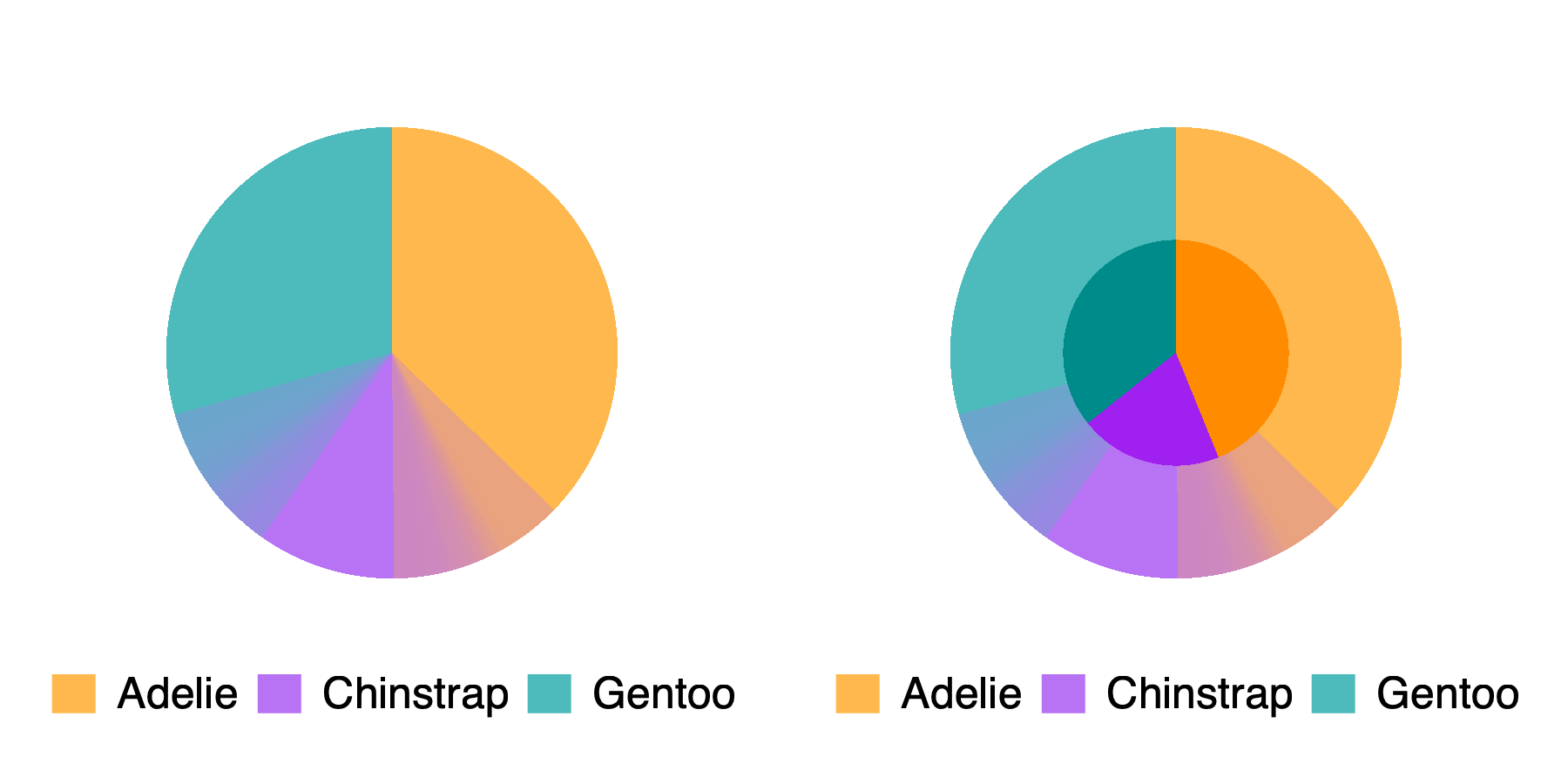}
\caption{When bootplot is applied to a base visualization of a pie chart, where the slices are uncertain, representing proportions of penguin species, it results in an ambiguated pie chart~\cite{olston2002visualizing} (left-hand side image). Adding a certain smaller pie chart overlay to the base visualization of a pie chart further emphasizes the uncertainty (right-hand side image). Both images are created with bootplot having $n=39$ samples.}
\label{fig:pie_chart_comparison}
\end{figure}

By distinguishing between certain elements (independent of the base distribution) and uncertain elements (dependent on the base distribution), we can better understand the types of visualizations that can be generated using the proposed approach. Examples in Figure~\ref{fig:bar_chart_comparison} and Figure~\ref{fig:pie_chart_comparison} further illustrate how certain and uncertain elements may appear in the foreground or background of a visualization.

In the extreme case where the base visualization contains only certain elements, the resulting aggregate image will be identical to every base visualization. In other words, the expressive power of our approach is at least as great as that of the tool used to create the base visualization. When uncertain elements are present, the aggregate image typically includes additional visual features that do not appear in any individual base visualization.

\subsection{Uncertain Text}\label{sec:fuzzy_text}

\begin{figure}[!t]
\centering
\includegraphics[width=3.5in]{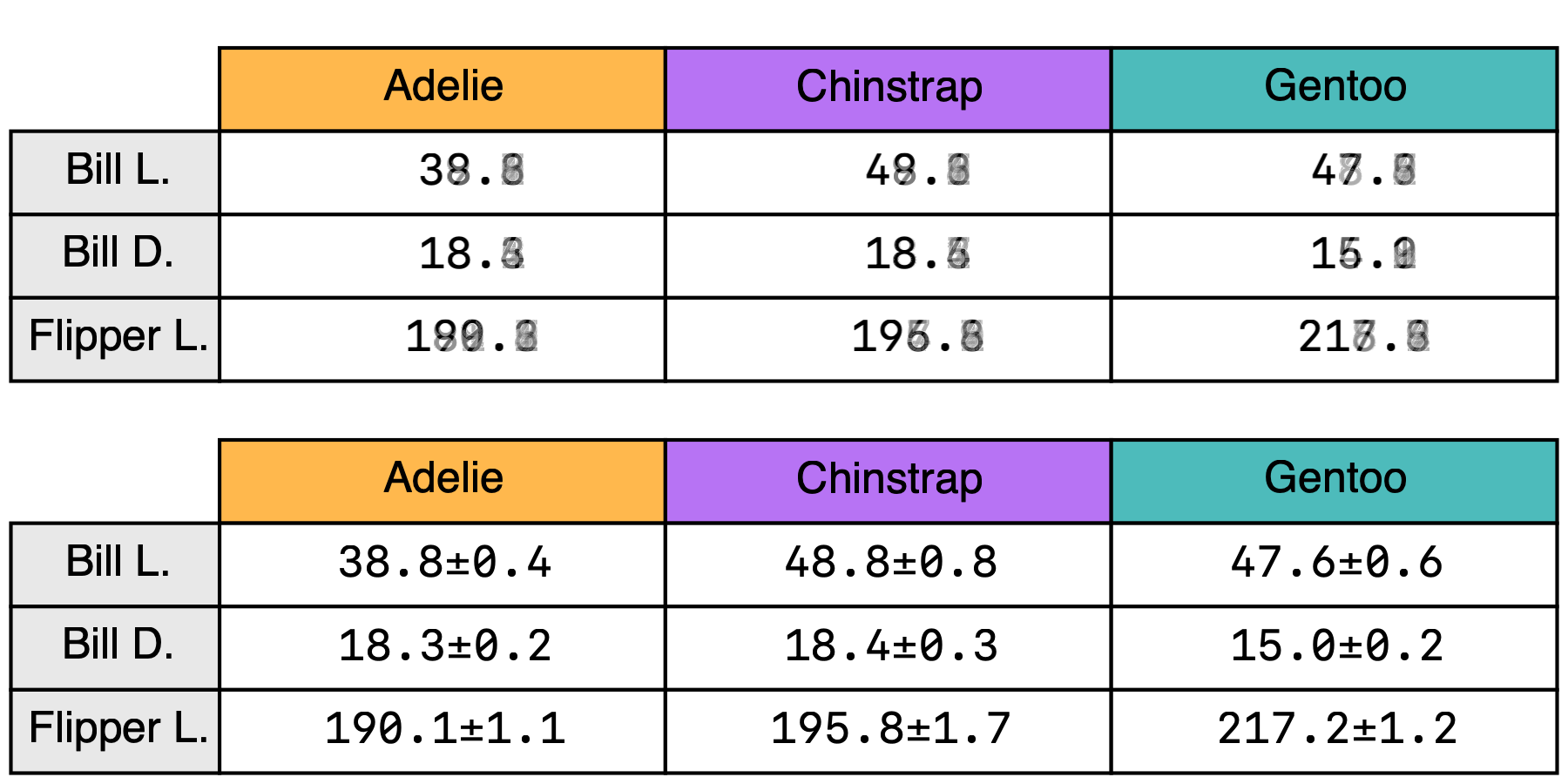}
\caption{Applying bootplot ($n=39$ samples) to a base visualization of a table listing mean estimates of different penguin features results in a blurry table (top image). For comparison, we include the standard approach of communicating uncertainty with standard errors (bottom image).}
\label{fig:table_comparison}
\end{figure}

Tables and text, when presented visually, also constitute static 2-D visualizations. Figure~\ref{fig:table_comparison} illustrates the application of bootplot to numerical tabular data and compares it to a standard table representing the same data. When text functions as an uncertain element, the resulting aggregate image conveys uncertainty through blurriness, as exemplified by the RMSE estimate in Figure~\ref{fig:examples:function}.

Because digits vary in shape, the exact boundaries of intervals may be difficult to discern. Nevertheless, the visualization still communicates the uncertainty associated with each digit: the more uncertain a digit, the less readable it becomes, potentially to the point of being unreadable.

This approach directly prevents the reader from interpreting uncertain digits, unlike standard methods that provide confidence intervals or standard errors. Traditional methods still rely on the user to correctly assess the magnitude of statistical error, often restricting interpretation to digits unlikely to be affected. For instance, in the top table of Figure~\ref{fig:table_comparison}, it is immediately apparent that the first decimal is largely noise and should likely not be reported if the goal is to infer the mean. In contrast, discerning this from the bottom table is more difficult.

Decreasing the discriminability of marks in a visualization as a means of communicating uncertainty or discouraging analysis in regions of high uncertainty is not new~\cite{gillmann2021uncertainty,maceachren2012visual,potter2011quantification,weiskopf2022uncertainty}. Visual features such as blur, transparency, and fuzziness naturally evoke perceptions of uncertainty and may be deliberately employed by designers~\cite{maceachren2012visual}. Practical applications include value-suppressing uncertainty palettes for bivariate maps, which allocate larger ranges of a visual channel to data when uncertainty is low~\cite{correll2018value}; texture-based strategies for visualizing uncertainty in time-dependent flows~\cite{botchen2005texture}; representing uncertainty in graphs~\cite{guo2015representing,wang2015ambiguityvis}; and sketchy or blurred trajectories~\cite{Boukhelifa2012}. However, the specific case of blurred and ambiguous text that naturally emerges from our approach is novel and merits further investigation.

A notable related work from outside uncertainty visualization is FatFonts~\cite{nacenta2012fatfonts}, a font for Arabic numerals in which the amount of ink (dark pixels) used to render each digit is proportional to its quantitative value. In our approach, more probable digits naturally appear more prominently. However, if a font is used in which each digit occupies the same amount of ink, the resulting ink distribution would reflect the relative frequency of each digit. This implementation would also require a font design with minimal overlap between digits.

\subsection{More Examples and a Summary of the Design Space}

Figure~\ref{fig:more_examples} presents additional visualizations generated using the proposed approach. Because uncertainty is encoded implicitly, emerging from the way information is represented in the base visualization, an exhaustive catalog of all possible outcomes is not feasible. To provide some insight, Table~\ref{tbl:design_space} summarizes the visual channels commonly used in statistical graphics and describes the corresponding implicit uncertainty representations that arise when each channel is employed in the base visualization.

We discuss several special and failure cases later in this section, but we first focus on situations in which the resulting visualization exhibits few or no issues. When representing a single uncertain quantity, visual channels such as position (x or y, or both), length, angle, area, and shape generally function without difficulty. Notable examples that extend beyond confidence intervals and bands include uncertain decision boundaries (see Figure~\ref{fig:ex:decision_boundary}), uncertain mean trajectories (see Figure~\ref{fig:ex:mean_traj}), uncertain PCA loadings (see Figure~\ref{fig:ex:pca}), and 2-D density regions for bivariate quantities of interest (see Figure~\ref{fig:ex:bivariate}).

For area (resulting in an uncertain region; see Figure~\ref{fig:ex:circular_area}) and shape (resulting in a mixture of shapes; see the individual icons in Figure~\ref{fig:ex:icon_chart}), it is more difficult to discern bounds for the uncertain variable. However, this difficulty primarily arises from the fact that area and shape are typically less suitable for encoding a univariate variable. Note that icon charts are commonly used to communicate risk in healthcare, and a visualization approach that replaces uncertain icons with question marks has been proposed~\cite{mcdowell2021communicating}.

The above claims also hold when there is more than one uncertain quantity, even when they overlap, provided they are independent. Examples include the regression curves in Figure~\ref{fig:polynomial_linear_comparison} and the radar plot in Figure~\ref{fig:ex:radar_chart}.

More complex behavior arises when multiple dependent quantities are present. A prototypical example is found in pie charts or stacked bar charts, where the size of one slice (or bar) affects the position of the others (see Figures~\ref{fig:pie_chart_comparison}, \ref{fig:ex:pie_chart}, and \ref{fig:ex:bar_chart}). This interdependence makes it more difficult to discern the uncertainty associated with elements between the two extreme, anchored elements. However, this issue is typically mitigated when the number of elements is small or when the chart is overlaid on all data in the base visualization. A related example is the Sankey flow diagram (see Figure~\ref{fig:ex:sankey}). Note that similar ambiguated pie and bar charts have been proposed~\cite{olston2002visualizing}, and a related gradient-based visualization has been introduced specifically for representing uncertainty in Sankey diagrams~\cite{VOSOUGH201944}.

Another type of interaction of visual channels can be observed when the base visualization is a bubble chart (see Figure~\ref{fig:ex:bubble}). In a bubble chart, a circle represents three variables, two with position and the third with area (radius, diameter). Therefore, both the position and the size of the bubble vary in the resulting uncertain bubble chart. The resulting uncertainty is more difficult to discern. This could be mitigated for position by plotting the center of each circle with a point, and for area by fixing the position on the data. This would result in an uncertain bubble chart that is a combination of the uncertain means in Figure \ref{fig:ex:bivariate} and uncertain areas in Figure~\ref{fig:ex:circular_area}.

The above observations highlight two considerations for bubble charts: (1) including a point at the center simplifies reading the position of observations, and (2) making the visual representation of one variable (area) dependent on other variables (position) can be problematic, as position may influence the interpretation of area. This again illustrates how our approach can reveal potential issues in the techniques employed in the base visualization.

Finally, note that for all of the cases discussed in this section, it is straightforward to interpret the two extremes with respect to the magnitude of uncertainty. When variability is very low, the result is practically identical to the base visualization and the relevant visual features are easy to discern. Conversely, when variability is high, uncertain visual features become impossible to distinguish, thereby not only conveying substantial uncertainty but also preventing interpretation of highly uncertain features.

\subsubsection{Color}

When colors (including transparency and texture) are used to represent an uncertain quantity in the base visualization, the resulting visualization will contain mixtures of those colors. When only two colors are used (typically for a binary uncertain quantity), there are no inherent color mapping issues. An implicit continuous color scale emerges, with the two colors at the extremes and the blends in between representing uncertainty (probability). Although not all values are directly discernible, there is a bijective mapping between color and the base distribution, so the uncertainty of every point in the resulting visualization is known. Examples include the waffle chart in Figure~\ref{fig:ex:waffle_chart} and the binary (trajectory hit/miss) heatmap in Figure~\ref{fig:ex:bin_traj}. The latter also illustrates how the implicit scale can be explicitly added as a certain element, since the two base colors are known. In practice, the discernibility of values in the resulting continuous color scale also depends on the choice of the two colors at the extremes, as in any color scale. In general, colors that are further apart in color space produce more perceptually distinguishable intermediate values.

When three or more colors are used (typically for three or more categories, or for a numerical variable represented with a gradient, transparency, or similar scale), there is an inherent issue: the mapping from the base distribution to colors is not injective. In other words, two different distributions can map to the same color in the aggregate image. For example, if a black-to-white grayscale gradient is used to represent a bounded continuous variable, a certain value of 0.5 (mid-gray) will appear identical to an uncertain value of 0.5, as long as the uncertainty is symmetric around 0.5 (an even mixture of lighter and darker grays).

When the number of colors is small and they are well separated, the issue is less problematic, particularly if the visualization can be anchored with a certain overlay or scale (see Figure~\ref{fig:ex:pie_chart} for an example of a pie chart with many low-frequency categories). However, when colors continuously overlap and span a gradient, this represents a fundamental limitation, preventing the discernment of uncertainty even in extreme cases of near-certainty or near-complete uncertainty.

This issue highlights not only a limitation of our approach but also an inherent difficulty in representing uncertainty when visualizations use color in this way. In statistical graphics, the use of more than two colors is typically reserved for representing values over 2-D continuous or discrete support (2-D density plots, heatmaps, or value encoding in correlograms). However, there are no straightforward methods to directly encode uncertainty in such visualizations. For example, existing techniques for visualizing uncertainty in heatmaps are limited to the addition of auxiliary visual elements \cite{chen2025relmap,holzhuter2012visualizing}.

\subsubsection{Non-Expressible Visualizations of Uncertainty}

Some standard visualizations of uncertainty cannot be expressed as the result of a sample-based overlay of uncertainty in base visualizations. A prototypical example is the one-dimensional (1-D) density plot. A 1-D density plot visualizes uncertainty in a univariate variable, but it cannot be generated by tying that variable to a visual channel and representing the uncertainty through overlay along that channel.

This type of non-expressibility does not imply that a density plot cannot be created directly with our approach. In fact, it can be trivially created from a certain base visualization of a density plot by taking only one sample. All visualizations can be created in this trivial way, but these non-expressible visualizations can only be created trivially, rendering our approach unnecessary.

\begin{figure*}[!p]
    \centering
    % Row 1
    \subfloat[circular area chart]{\includegraphics[width=0.22\linewidth]{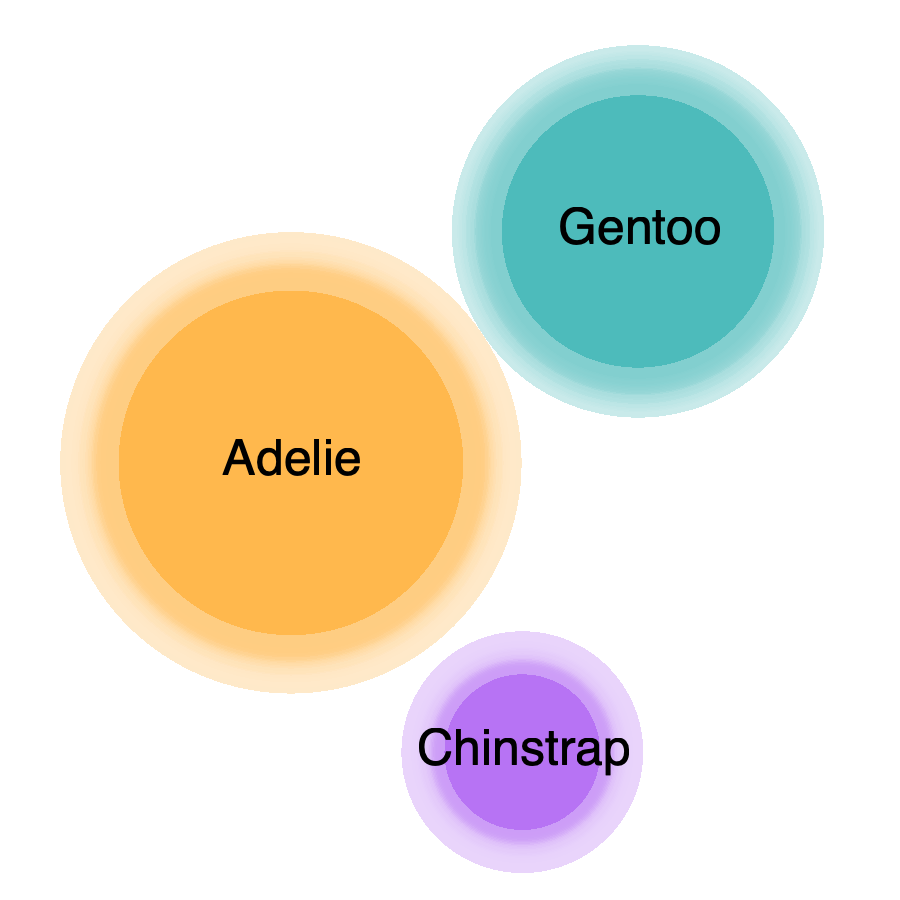}\label{fig:ex:circular_area}}
    \hfill
    \subfloat[decision boundary]{\includegraphics[width=0.25\linewidth]{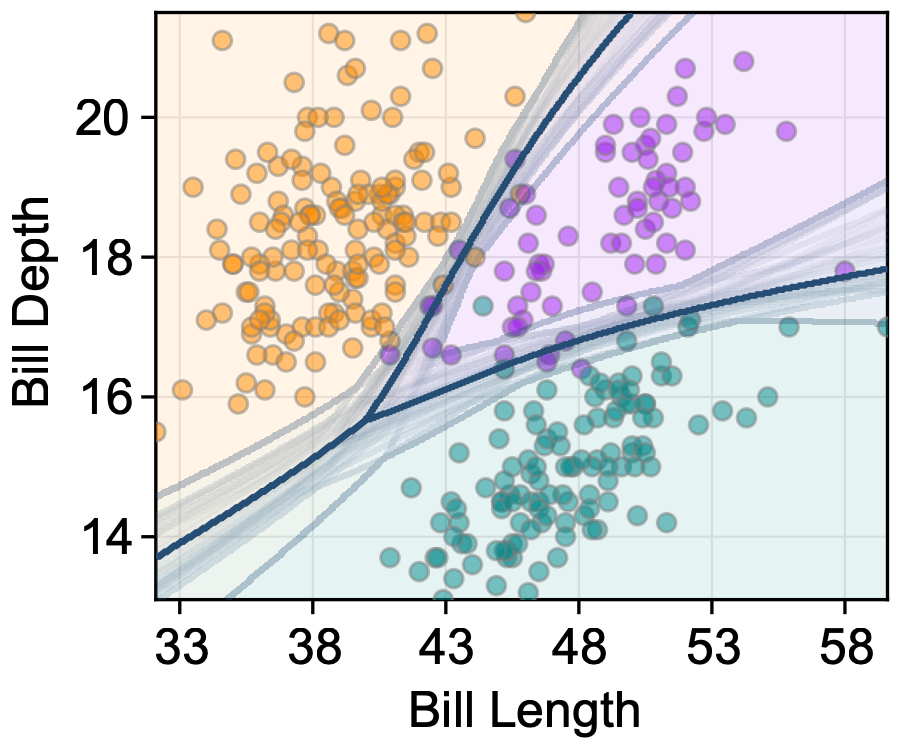}\label{fig:ex:decision_boundary}}
    \hfill 
    \subfloat[waffle chart]{\includegraphics[width=0.21\linewidth]{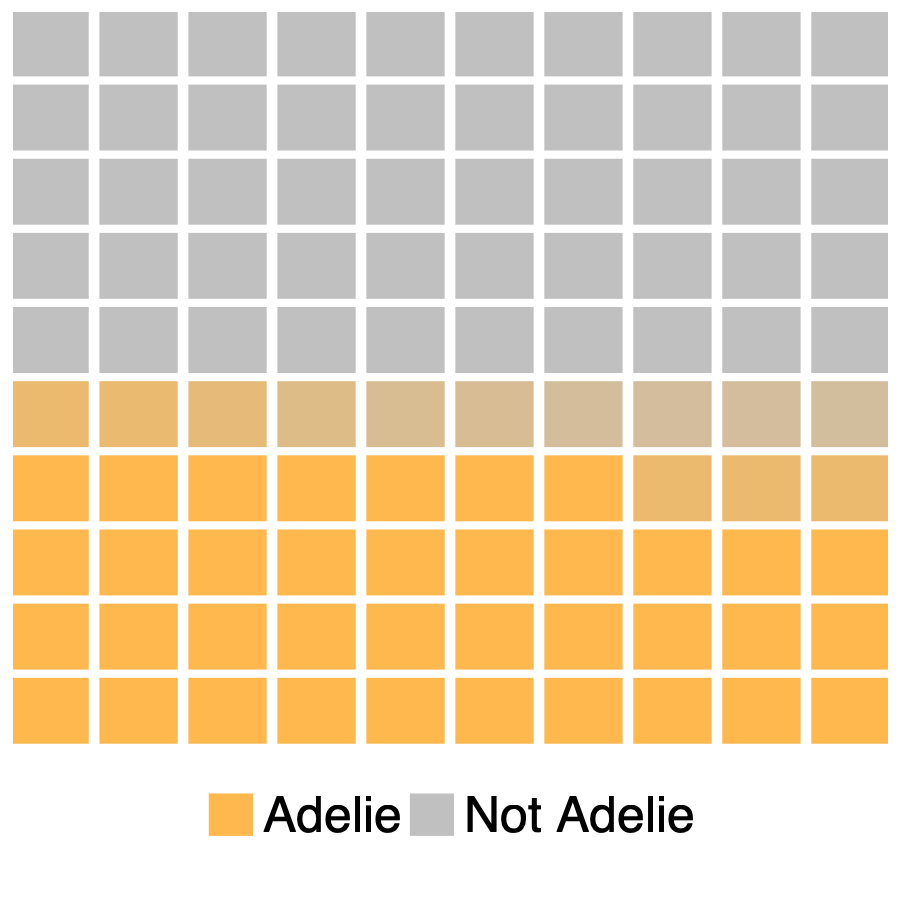}\label{fig:ex:waffle_chart}}    
    \hfill
    \subfloat[icon chart]{\includegraphics[width=0.195\linewidth]{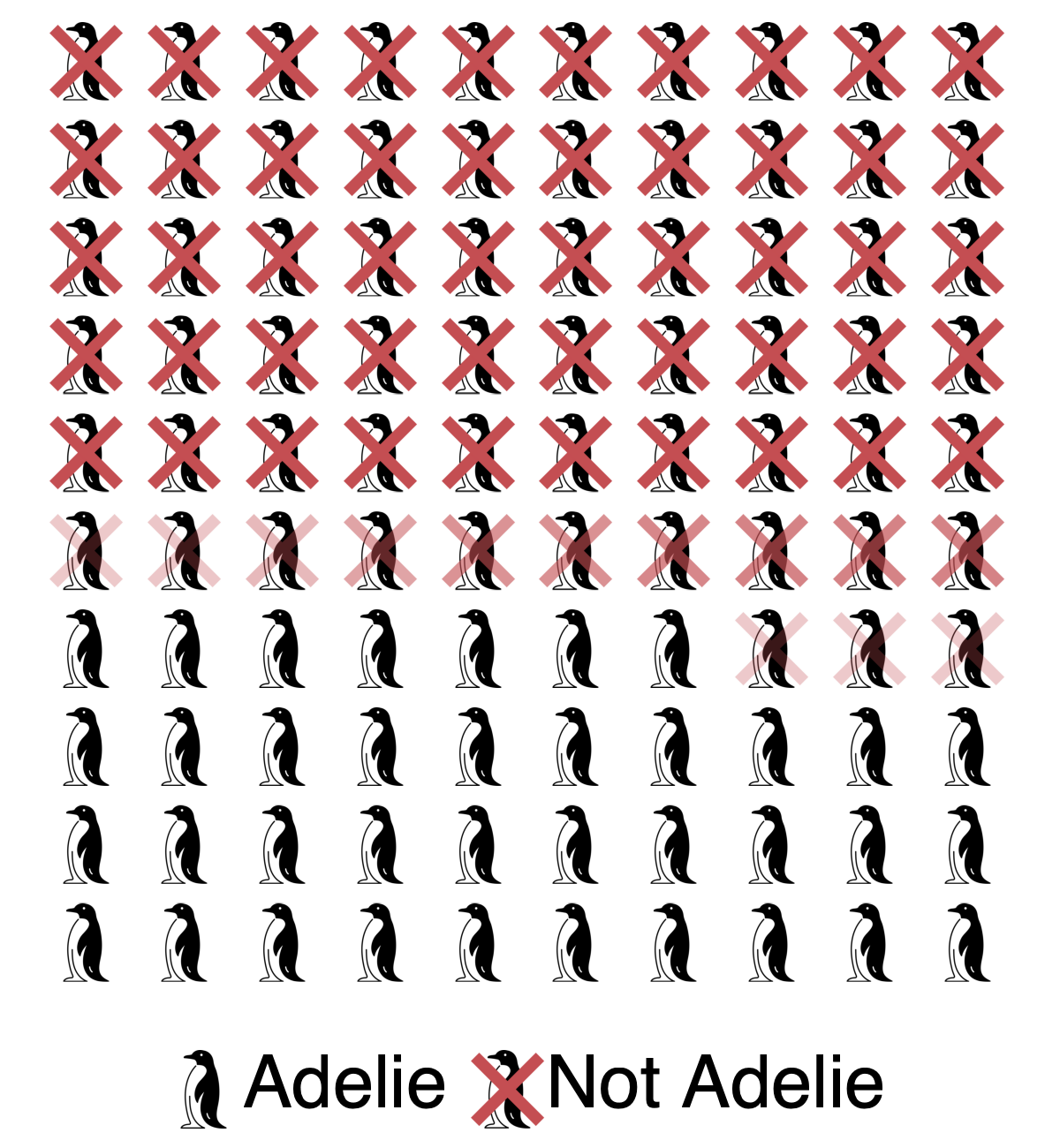}\label{fig:ex:icon_chart}}\\
    % Row 2
    \subfloat[density plot]{\includegraphics[width=0.25\linewidth]{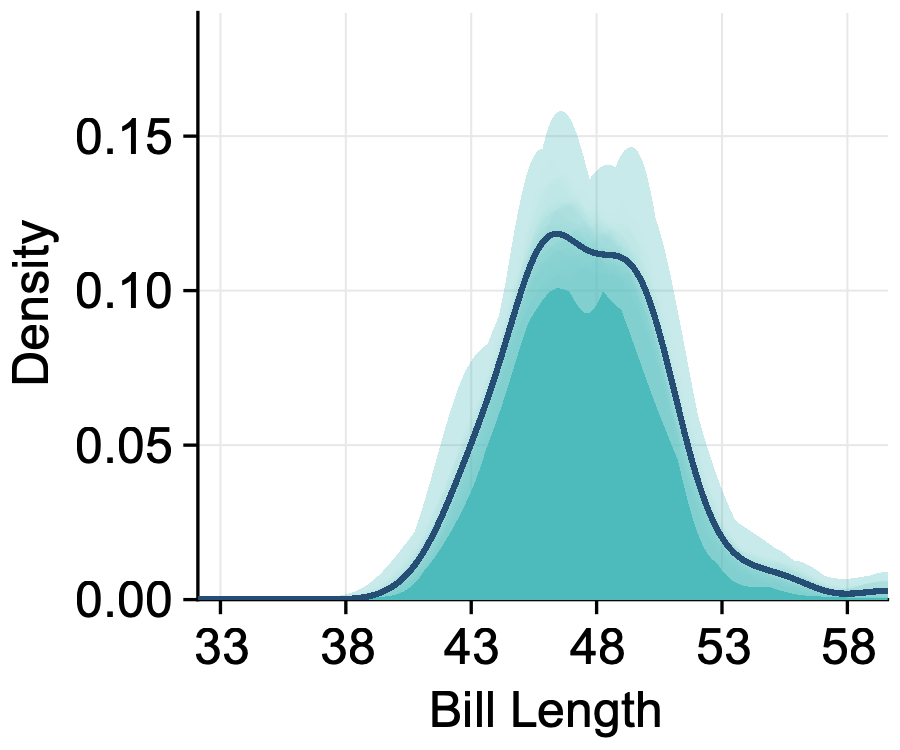}\label{fig:ex:density_plot}}
    \hfill
    \subfloat[radar plot]{\includegraphics[width=0.22\linewidth]{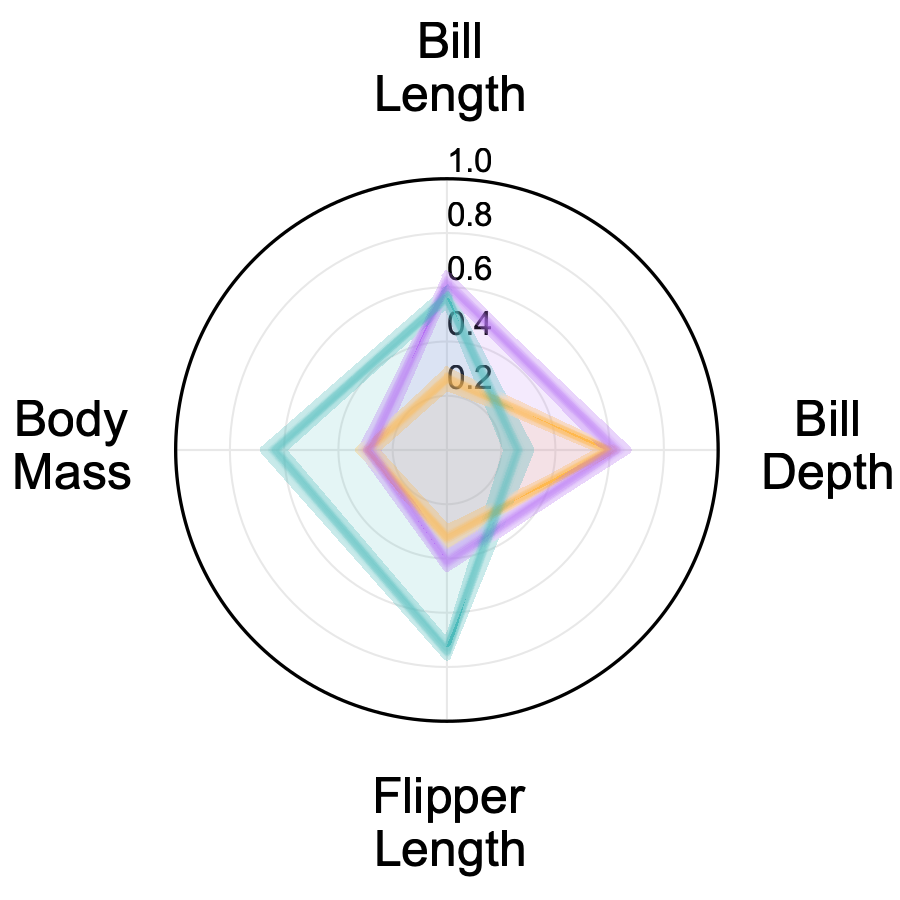}\label{fig:ex:radar_chart}}
    \hfill
    % Last column: stacked images
    \begin{minipage}[b]{0.44\linewidth}
        \centering
        \subfloat[horizontal stacked bar chart]{\includegraphics[height=0.12\linewidth,trim=0 0 0 220, clip]{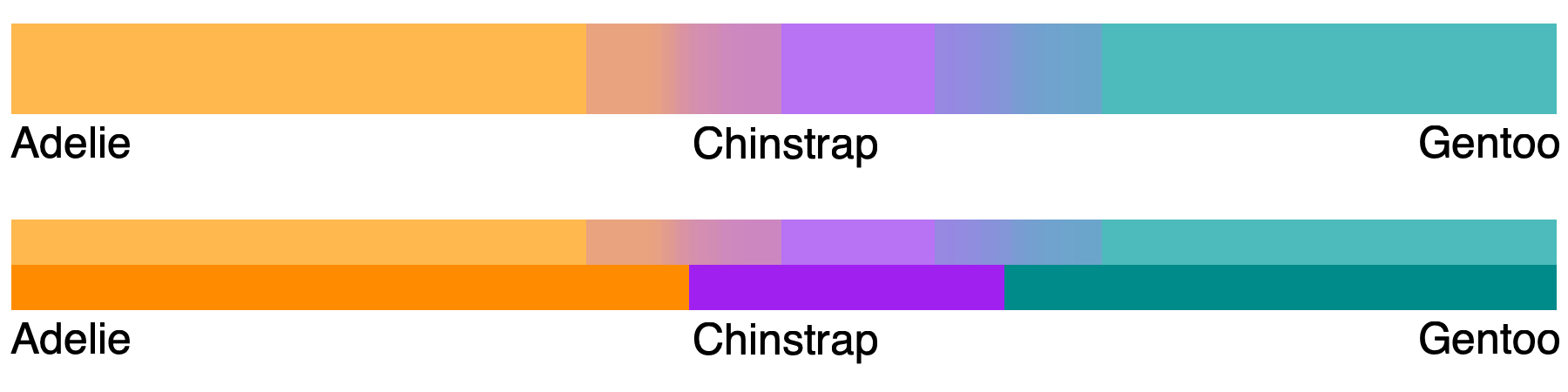}\label{fig:ex:bar_chart}}\\
        \subfloat[Sankey diagram]{\includegraphics[height=0.25\linewidth]{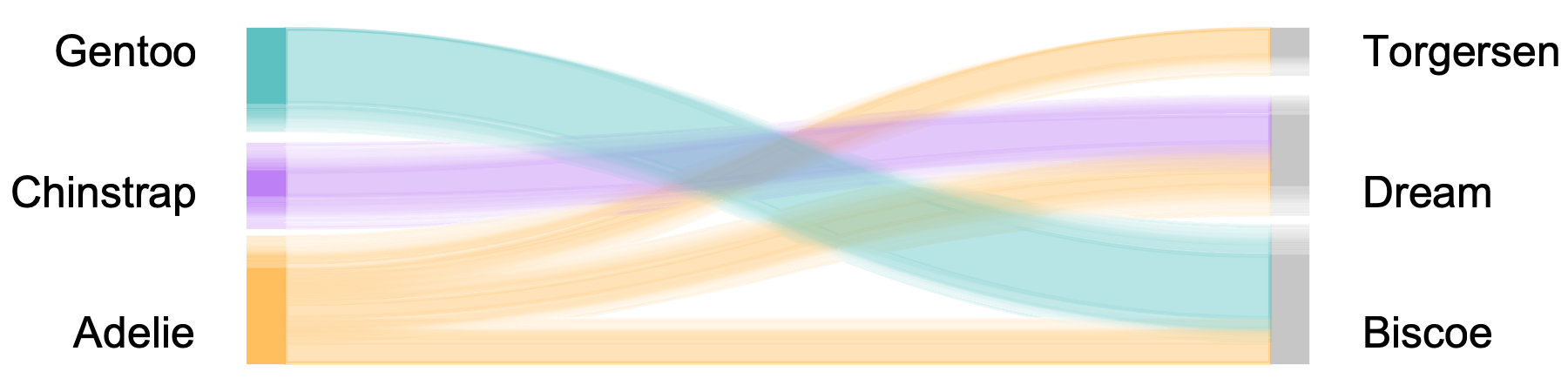}\label{fig:ex:sankey}}
    \end{minipage}
    \\ 
    % Row 3
    \subfloat[pie chart]{\includegraphics[width=0.4\linewidth,trim=0 100 0 100, clip]{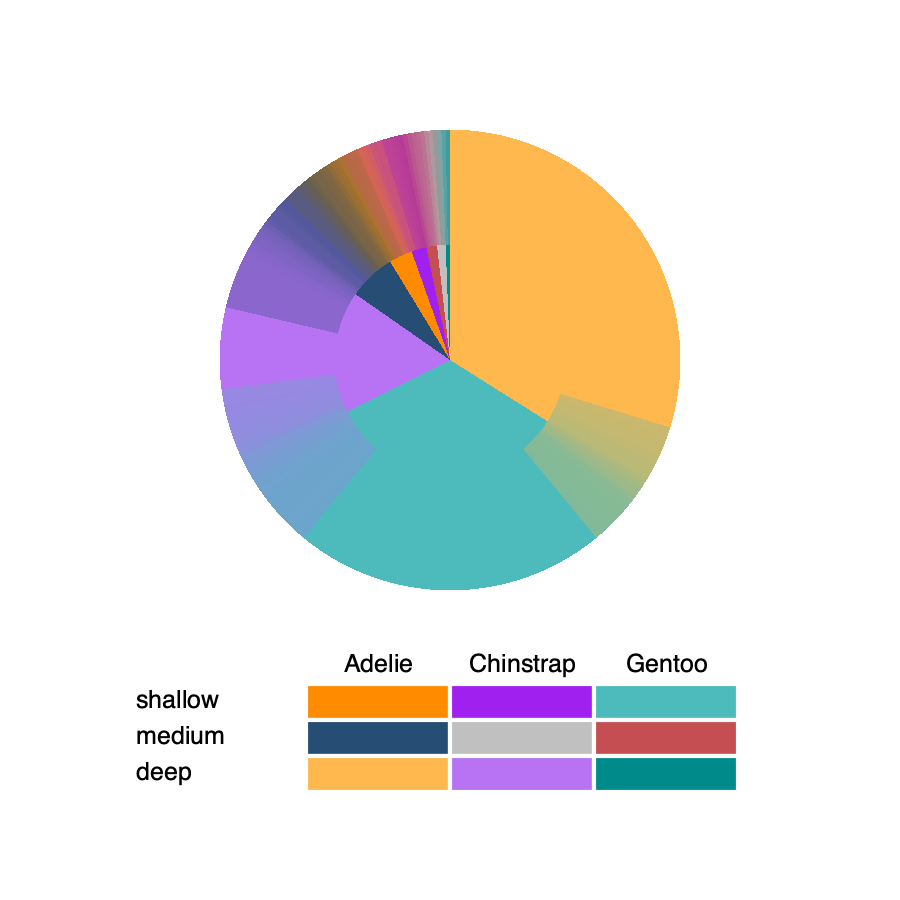}\label{fig:ex:pie_chart}}
    \hfill  
    \subfloat[mean trajectory]{\includegraphics[width=0.3\linewidth,trim=0 0 0 0, clip]{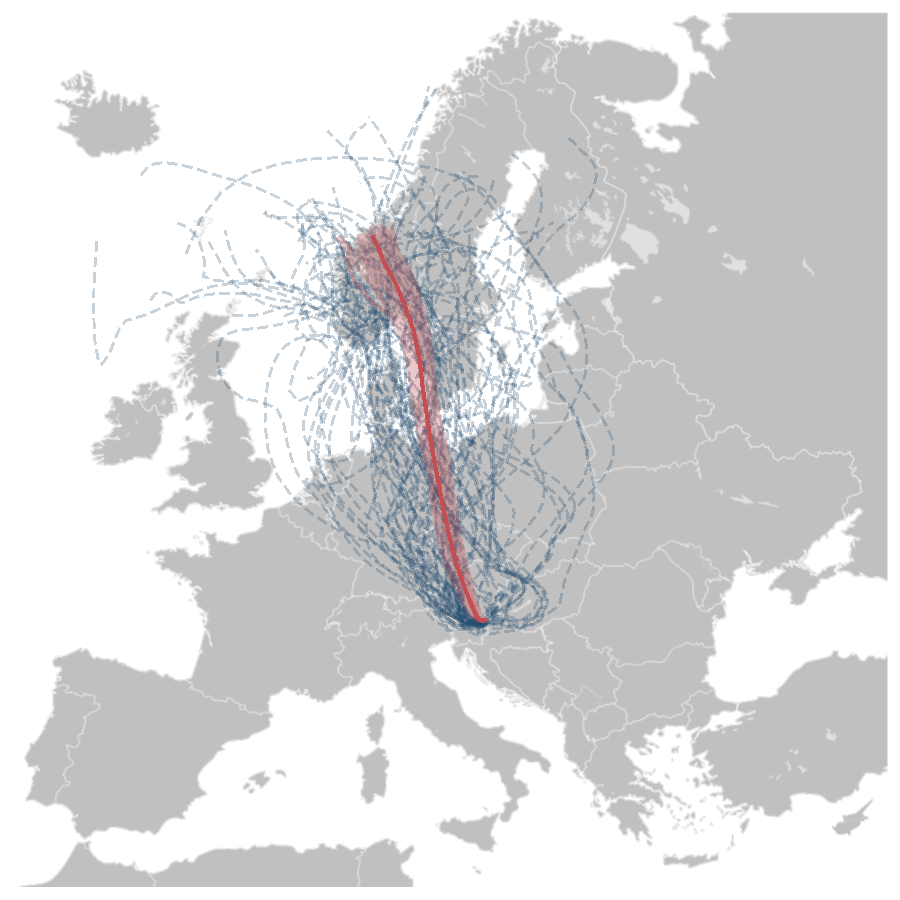}\label{fig:ex:mean_traj}}
    \hfill
    \subfloat[binary heatmap]{\includegraphics[width=0.3\linewidth,trim=0 0 0 0, clip]{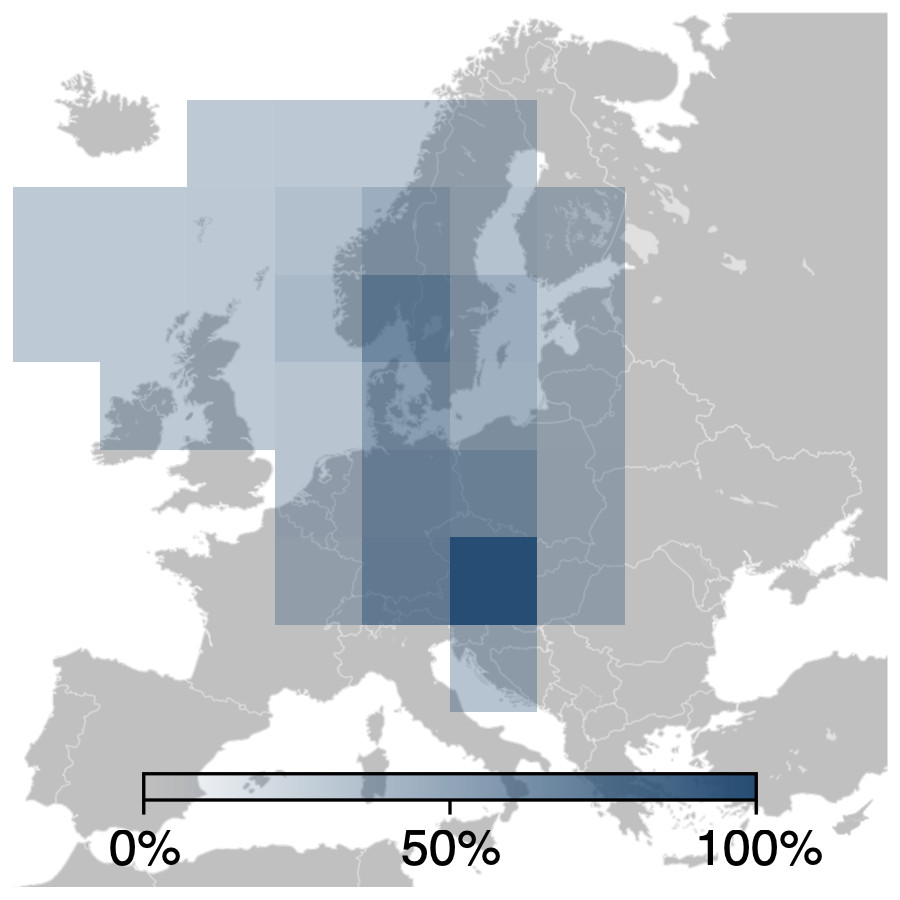}\label{fig:ex:bin_traj}}
    \\   
    % Row 4
    \subfloat[PCA loadings]{\includegraphics[width=0.3\linewidth]{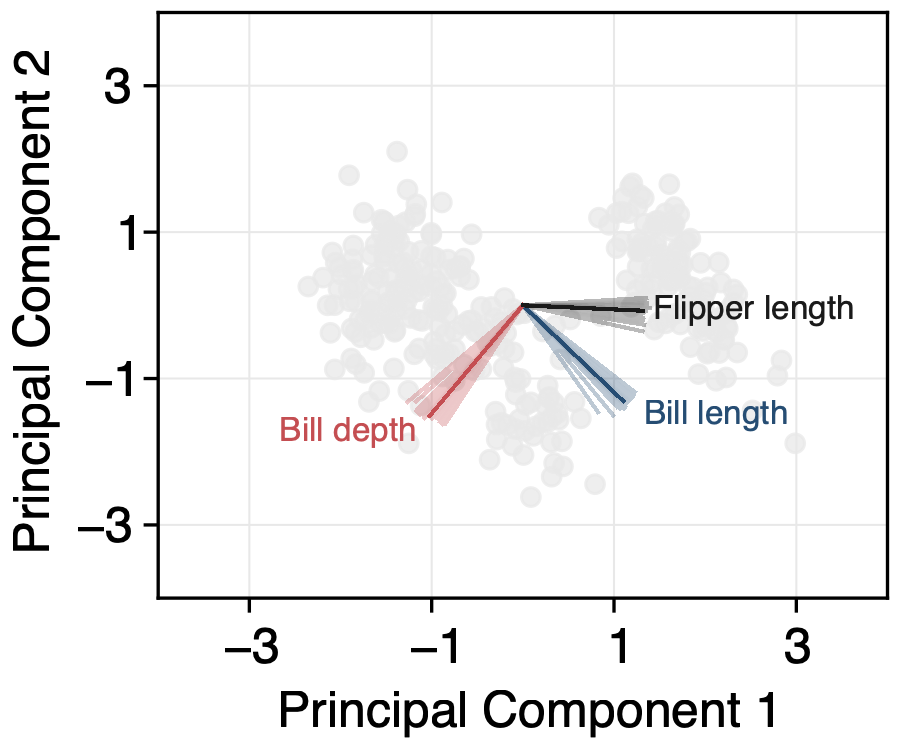}\label{fig:ex:pca}}
    \hfill
    \subfloat[bivariate means]{\includegraphics[width=0.3\linewidth]{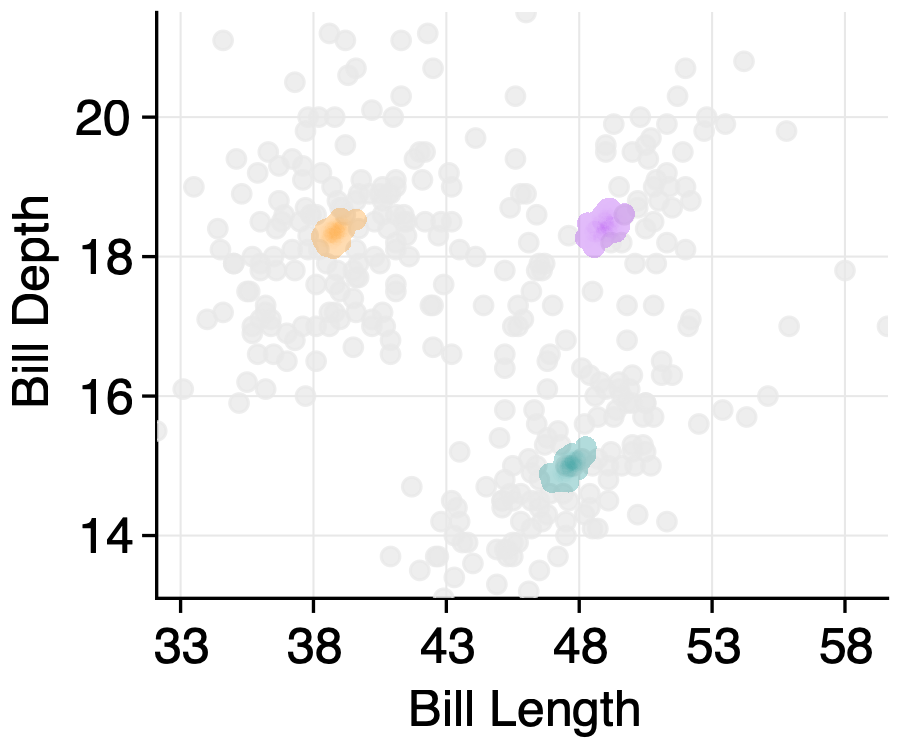}\label{fig:ex:bivariate}}
    \hfill    
    \subfloat[bubble chart]{\includegraphics[width=0.28\linewidth]{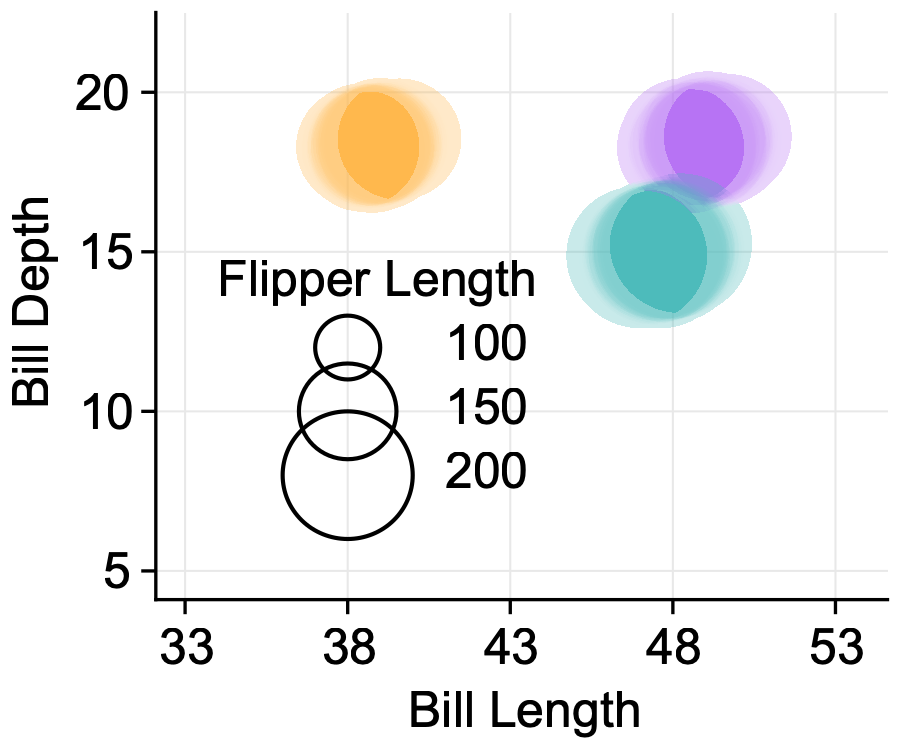}\label{fig:ex:bubble}}
    \caption{More examples of visualizations of uncertainty created by applying the proposed approach to the base visualization. The base visualization used in each case is indicated below the corresponding example, and the number of samples used by the bootplot library is $n=39$. }
    \label{fig:more_examples}
\end{figure*}

There are two important cases where our approach remains useful even for non-expressible visualizations. First, when other uncertain elements are present in the plot (for example, an estimate of the mean), and second, when an uncertain density plot is desired. In the latter case, when the density plot represents an empirical distribution, it is itself an uncertain estimate of the underlying true distribution. While a standard density plot is non-expressible in our framework, an uncertain density plot can be represented as a mixture of areas or shapes (see Figure~\ref{fig:ex:density_plot}).

Other common visualizations of uncertainty fall into the same category as density plots, including contour plots, violin plots, quantile dot plots, histograms, and boxplots. Visualizations based on displaying all samples are also non-expressible. However, it is generally unnecessary to visualize uncertainty in these cases, since there is either no variability (for example, scatterplots) or variability is introduced solely to improve visibility (for example, jitter plots, beeswarm plots).

\begin{table*}[!t]
\centering
\caption{A summary of how uncertainty in different base visualization channels implicitly creates uncertainty visual encodings.}
\label{tbl:design_space}
\begin{tabular}{|p{0.18\textwidth}|p{0.25\textwidth}|p{0.48\textwidth}|}
\hline
\textbf{base visualization channel} & \textbf{result (representation of uncertainty)} & \textbf{examples} \\
\hline
position (x or y) & interval (length); density ribbon (length, opacity) & Basic examples in Figures~\ref{fig:bootplot} and \ref{fig:parameters_choice}. \newline Confidence intervals in Figure~\ref{fig:points_ci_comparison} and right-hand side image in Figure~\ref{fig:bar_chart_comparison}. \newline The confidence bands in Figure~\ref{fig:polynomial_linear_comparison} are composed of multiple intervals.\\
\hline
position (x and y) & area (area); 2-D density (area, opacity)& Bivariate means in Figure~\ref{fig:ex:bivariate}. \newline
   Mean trajectory line results in an area in Figure~\ref{fig:ex:mean_traj}. \newline
  Decision boundary lines result in an area in Figure~\ref{fig:ex:decision_boundary}. \newline      
   Radar plot in Figure~\ref{fig:ex:radar_chart}. \newline
   Bubble chart in Figure~\ref{fig:ex:bubble}.
\\
\hline
length & (half) density ribbon (length, opacity) &
Density strip bars in the left-hand side image in Figure~\ref{fig:bar_chart_comparison}. \newline
Stacked bar chart in Figure~\ref{fig:ex:bar_chart}. \newline
Uncertain Sankey diagram in Figure~\ref{fig:ex:sankey}.\newline
 Density plot in Figure~\ref{fig:ex:density_plot}, if we consider the heights.    
\\
\hline
angle & directional cone (area, opacity) &
  Loadings in Figure~\ref{fig:ex:pca}. \newline
   Pie charts in Figures~\ref{fig:pie_chart_comparison} and \ref{fig:ex:pie_chart}.
\\  
\hline
area & density region (area, opacity) &
    Circular area chart with fixed center in Figure~\ref{fig:ex:circular_area}. \newline
   Radar plot in Figure~\ref{fig:ex:radar_chart} if we consider the area as a whole. \newline
   Size of bubbles in Figure~\ref{fig:ex:bubble}.
\\
\hline
shape &  mixture of shapes (shape, blur/focus) &
   Blurry text for RMSE in Figure~\ref{fig:polynomial_linear_comparison} and in the top table in Figure~\ref{fig:table_comparison}. \newline
   Penguin icons in the icon chart in Figure~\ref{fig:ex:icon_chart}. \newline
   Density plot in Figure~\ref{fig:ex:density_plot}. \\
\hline
color (2 colors) &  mixture of colors (color) & 
    Each individual square in the waffle chart in Figure~\ref{fig:ex:waffle_chart}. \newline 
   Pie charts in Figures~\ref{fig:pie_chart_comparison} and \ref{fig:ex:pie_chart}. \newline
   Stacked bar chart in Figure~\ref{fig:ex:bar_chart}.  \\

\hline
color (3 or more) & mixture of colors (color)& Pie chart in Figure~\ref{fig:ex:pie_chart}.  \\
\hline
\end{tabular}
\end{table*}

\subsection{Visual Inference}\label{sec:examples_inference}

We derive coverage guarantees for intervals, which extend to point-wise bands, and inference for predetermined regions of the visualization, allowing our approach to support visual inference. The implied intervals for point estimates in our examples correspond to CIs, providing a direct measure of uncertainty. Through the duality between CIs and hypothesis tests, these intervals can be used for null-hypothesis significance testing. For example, suppose that we are interested in testing the hypothesis that the Adelie penguin expected bill length is 40 (see Figure~\ref{fig:points_ci_comparison}). Because the interval does not overlap that value, we can reject the hypothesis with at least 2.5\% confidence. Note that two-sided intervals can also be used to construct one-sided bounds.

For visual inference, the method inherits the properties of the underlying procedure used to generate the base distribution. In our examples, non-parametric resampling is used to produce percentile intervals, which are appropriate in most cases when the data set is sufficiently large~\cite{Zrimvsek2025}. For more critical inference tasks, the base distribution could be obtained using a more sophisticated method, or visual inference might not be applied at all. The interpretation also depends on the inference framework. In a Bayesian setting, posterior samples would yield credible intervals and posterior probabilities rather than frequentist confidence intervals.

We do not provide guarantees for general bivariate distributions, which include special cases such as convex hull coverage and simultaneous confidence bands. These require knowledge of the dependency structure among quantities of iterest, and deriving coverage even under known distributions is challenging. See, for example, \cite{han2014simultaneous} for a review of simultaneous confidence bands for linear models, and \cite{son2015new} for derivations of convex hull coverage for some simple distributions. Establishing guarantees under weaker assumptions is an open problem and would constitute a substantial contribution to probability and statistics, beyond the scope of this work.

While general inference for regions observed in the image post-hoc is challenging, the inference for a predetermined region, as we show in Section~\ref{sec:coverage}, is relatively simple. For example, suppose we are interested in testing the hypothesis that the air mass trajectory will go over Western France (see Figure~\ref{fig:ex:bin_traj}). If this region is determined before creating the visualization and because none of the 39 samples go over Western France, we can reject the hypothesis with at least 95\% confidence.  

It is worth noting that inference over more than one variable at a time is uncommon in practice, both because constructing confidence areas (for example, confidence ellipses) is challenging and because interpreting them is difficult. Spatial data analysis, such as in our trajectory example, is one of the few scenarios where confidence areas are both naturally relevant and relatively easy to interpret. For these reasons, we do not consider the limitation to variation along a single axis or predetermined regions to be a substantial impediment to practical applications.

\section{Discussion}\label{sec:discussion}

\subsection{Relationship with Related Work}\label{sec:related}

The use of location and blur to communicate uncertainty is not new, and several visualizations produced by our approach resemble previously explored techniques. Furthermore, any visualization generated with our method, including novel forms, can also be implemented conventionally by explicitly encoding visual elements and specifying their locations, transparency, and other relevant properties. The key distinction of our approach is in the process used to create the visualization rather than in the visualization itself.

Similarly, the use of confidence intervals for quantifying uncertainty and supporting visual inference is well established. What is novel in our approach is that uncertainty is not explicitly quantified or directly encoded. Only the base visualization and underlying data are used. Despite this generalization, we maintain theoretical coverage guarantees, providing a new perspective on the relationship between conventional uncertainty quantification and sample-based visualization methods. A related idea are lineup tests~\cite{hofmann2012graphical,wickham2010graphical}. However, in lineup tests, the actual data visualization is shown in a lineup with several visualizations from a null-hypothesis and the identification rate for one or across more users is used to test the null-hypothesis. In our case, the samples either construct a rejection region, or their (lack of) appearance in a predetermined region is used to test a hypothesis.

Sample-based approaches are closely related, as they also avoid explicit quantification of uncertainty, but they still require the encoding and visualization of multiple samples. Our work extends these approaches in two ways. First, we demonstrate that explicit encoding of samples is not necessary. Second, we provide theoretical coverage guarantees that are not offered by traditional sample-based methods. This positions our approach as a generalization of sample-based overlay techniques.

An additional advantage of our approach is practical and emerges when considering multiple scenarios rather than a single one. Without our method, each scenario would require a scenario-specific base visualization along with explicit encoding and visualization of multiple samples. In contrast, our approach only requires a different base visualization for each scenario, simplifying the conceptual framework and reducing both code complexity and maintenance. 

Our approach complements general visualization tools (Matplotlib, ggplot2~\cite{Wickham2016Ggplot2}, etc.), since users need familiarity with at least one such tool to flexibly create base visualizations. More specialized tools for visualizing uncertainty, whether they provide wrappers to simplify uncertainty visualization (Seaborn) or delve deeper into its structure (ggdist~\cite{Kay2024}, PGoG \cite{pu2020probabilistic}), still require explicit quantification and visual encoding of uncertainty. While these approaches offer valuable insights, our approach is fundamentally different. 

Recent work by Mason et al.~\cite{mason2026noisy} and the ggdibbler package~\cite{mason2026ggdibbler} similarly argues for supplying distributions to visualizations, leveraging the resulting noise to perceptually suppress weak signals. However, they limit their approach to the Grammar of Graphics (ggplot2 in particular), forcing the user to explicitly manage distributions and overlay rules. Our work diverges theoretically by operating entirely in the image space, treating the base visualization as a black-box statistical functional. This allows our approach to naturally preserve complex joint distributions, apply to arbitrary visual marks, and provide formal mathematical coverage guarantees for many of the resulting visualizations.

Furthermore, if versatility is important, then these tools support only a limited set of uncertainty visualizations. For the rest, the user must both quantify the uncertainty and design its visual encoding manually in the coding pipeline. To provide some insight, Table~\ref{tab:practical_versatility} summarizes the versatility of the above mentioned tools for producing uncertainty visualization examples from Section~\ref{sec:examples} that were produced by bootplot. We exclude PGoG from the table because it is a prototype.

\begin{table*}[!t]
\centering
\caption{Practical versatility of common visualization tools for reproducing the uncertainty visualization examples from Section~\ref{sec:examples} that were produced by bootplot. ``$\checkmark$'' indicates the existence of a wrapper that quantifies uncertainty and designs its visual encoding, including overlay where relevant. ``$\circ$'' means the user must quantify uncertainty, but the wrapper provides the visual encoding by default. ``$\times$'' indicates that the user must explicitly quantify uncertainty and design the visual encoding themselves, including overlay choices when relevant. Grey rows indicate uncertainty visualizations where bootplot is the only tool providing such practical versatility.}
\label{tab:practical_versatility}
\begin{tabular}{|p{0.32\textwidth}|>{\centering\arraybackslash}p{0.105\textwidth}>{\centering\arraybackslash}p{0.105\textwidth}>{\centering\arraybackslash}p{0.105\textwidth}>{\centering\arraybackslash}p{0.105\textwidth}>{\centering\arraybackslash}p{0.105\textwidth}|}
\hline
\textbf{visualization} & \textbf{matplotlib} & \textbf{ggplot2} & \textbf{ggdist} & \textbf{seaborn} & \textbf{ggdibbler\textsuperscript{*}} \\
\hline
Confidence interval for mean & $\circ$ & $\checkmark$ & $\circ$ & $\checkmark$ & $\times$ \\
\hline
Point-wise confidence band around regression line & $\circ$ & $\checkmark$ & $\circ$ & $\checkmark$ & $\times$ \\
\hline
Bar chart with confidence interval for mean & $\circ$ & $\checkmark$ & $\circ$ & $\checkmark$ & $\times$ \\
\hline
\rowcolor{rowgray}
Ambiguous pie chart & $\times$ & $\times$ & $\times$ & $\times$ & $\times$ \\
\hline
\rowcolor{rowgray}
Ambiguous bar chart & $\times$ & $\times$ & $\times$ & $\times$ & $\times$ \\
\hline
\rowcolor{rowgray}
Overlaid numbers & $\times$ & $\times$ & $\times$ & $\times$ & $\times$ \\
\hline
\rowcolor{rowgray}
Uncertain circular area chart & $\times$ & $\times$ & $\times$ & $\times$ & $\times$ \\
\hline
\rowcolor{rowgray}
Uncertain decision boundary & $\times$ & $\times$ & $\times$ & $\times$ & $\times$ \\
\hline
\rowcolor{rowgray}
Uncertain waffle chart & $\times$ & $\times$ & $\times$ & $\times$ & $\times$ \\
\hline
\rowcolor{rowgray}
Uncertain icon chart & $\times$ & $\times$ & $\times$ & $\times$ & $\times$ \\
\hline
\rowcolor{rowgray}
Uncertain radar plot & $\times$ & $\times$ & $\times$ & $\times$ & $\times$ \\
\hline
\rowcolor{rowgray}
Uncertain density plot & $\times$ & $\times$ & $\times$ & $\times$ & $\times$ \\
\hline
\rowcolor{rowgray}
Band around mean trajectory & $\times$ & $\times$ & $\times$ & $\times$ & $\times$ \\
\hline
Uncertain binary heat map & $\circ$ & $\circ$ & $\times$ & $\circ$ & $\times$ \\
\hline
\rowcolor{rowgray}
Uncertain Sankey diagram & $\times$ & $\times$ & $\times$ & $\times$ & $\times$ \\
\hline
\rowcolor{rowgray}
Uncertain PCA vectors & $\times$ & $\times$ & $\times$ & $\times$ & $\times$ \\
\hline
Uncertain bivariate means & $\circ$ & $\circ$ & $\times$ & $\circ$ & $\times$ \\
\hline
\rowcolor{rowgray}
Uncertain bubble chart & $\times$ & $\times$ & $\times$ & $\times$ & $\times$ \\
\hline
\end{tabular}

\vspace{0.3em}

\begin{minipage}{0.99\textwidth}
\footnotesize
\textsuperscript{*} To use ggdibbler, the user must first quantify uncertainty to be able to provide an uncertainty distribution, so we cannot mark it ``$\checkmark$'' by design. Although ggdibbler provides useful functionality beyond standard ggplot2, we mark it with ``$\times$'' rather than ``$\circ$'' because the user must still explicitly manage the overlay rules needed to produce these visualizations.
\end{minipage}

\end{table*}

\subsection{Who are the Target Users?}

Our approach requires only basic coding skills and the ability to create base visualizations (without uncertainty), either through custom code or using libraries such as Matplotlib or ggplot2. For users already familiar with tools like Seaborn, ggdist, or similar packages, our method can serve as a complement: a base visualization created in Seaborn can be extended to visualize uncertainty in scenarios not directly supported by these tools.

We hypothesize that the primary application of our approach is as a teaching tool in introductory courses on applied statistics. Visualizing uncertainty and understanding how it is generated can facilitate a faster and deeper comprehension of underlying statistical concepts. Additionally, we anticipate that the method would be a valuable addition to the toolkit of quantitative data analysis professionals, enabling rapid prototyping of uncertainty visualizations for both standard and non-standard scenarios.

\subsection{Computational Complexity and Runtime}

Our approach consists of three steps: obtaining the base distribution, applying the base visualization to $n$ samples, and aggregating the resulting images. Since any alternative method must also obtain the base distribution and produce an image, the only additional factor is $n$. In practice, $n$ is fixed based on the desired coverage and does not grow with problem size. Asymptotically, aggregation of the images and repeated base visualizations dominate computation, requiring $O(n \cdot K + n \cdot C_\text{base})$ operations, where $K$ is the number of pixels and $C_\text{base}$ is the cost of a single visualization. In most practical examples shown above, base visualizations take approximately tenths of a second and, with parallelization of the image aggregation, making aggregate images computable in approximately 1 second. For larger images, such as those shown in the supplementary material, image aggregation remains parallelized, and the total runtime is still approximately $n$ times the cost of rendering a single base visualization. Improvements in hardware will further reduce any runtime differences compared to faster alternatives.

\section{Conclusion}

This work introduces a general approach for visualizing uncertainty that treats statistical graphics as random images generated from a base distribution of the quantities of interest. By requiring only a base distribution and a base visualization, the method produces uncertainty encodings implicitly through pixel-level aggregation, while providing formal coverage guarantees for many common visual inference tasks, for example, null hypothesis significance testing, interval based inference, and inference for predetermined regions of the visualization.

The approach generalizes sample-based overlay methods and demonstrates how uncertainty intervals and bands emerge naturally from samples, without the need for explicit quantification or dedicated visualization routines. This establishes an important connection between sample-based approaches and standard non-visual inference techniques. And, the approach is a useful technique for identifying poor statistical graphics practices, flaws in the base visualization, and which types of visualizations are more amenable to visualization of uncertainty.

A reference implementation, bootplot, although currently limited to IID cases, illustrates the versatility of the approach and is already a useful substitute or complement to existing tools. It can generate uncertainty visualizations for all common scenarios as well as many non-standard cases, including entirely novel representations. The reference implementation can also be viewed as a generalization of the non-parametric bootstrap to image functionals.

\subsection{Directions for Future Work}\label{subsec:futurework}

We leave the exploration of coverage guarantees for convex hulls and simultaneous confidence bands to future work.

This study focused exclusively on static 2-D visualizations of uncertainty. Nevertheless, because interactive visualizations, animations, and videos can be viewed as sequences of 2-D representations, our approach may be extendable to non-static settings.

Several important questions remain that require user evaluation. In particular, it is unclear (1) whether this general approach would serve as an effective teaching or practical tool, (2) how valuable, interpretable, and comparable in practice the novel visualization techniques it produces, such as blurry text, are, and (3) whether users can still meaningfully aggregate visually separated samples into intervals or bands when these are only implied, especially given that formal coverage guarantees do not by themselves imply perceptual readability or effective decision support.

From the perspective of the bootplot reference implementation, it is also important to investigate how it can be adapted to more complex scenarios, such as hierarchical sampling schemes, while preserving simplicity and usability.

\section*{Acknowledgments}

This work was partially funded by the Slovenian Research Agency Grants No. P2-0442 and J5-60084. The authors would also like to thank the anonymous reviewers of an earlier version of this paper for their constructive comments, which have helped substantially improve the quality of this work.

\bibliography{bibliography}
\bibliographystyle{plain}

\end{document}